\documentclass{article}  
\usepackage{environ}
\usepackage[
a4paper,
inner=1.5cm,
outer=1.5cm,
top=2.5cm,
bottom=2.5cm
]{geometry}

\usepackage[utf8]{inputenc}
\usepackage{amsmath, amssymb, bm, nicefrac}
\usepackage{graphicx}
\usepackage{hyperref}
\usepackage{url}
\usepackage{float}
\usepackage{authblk}
\usepackage{xcolor}

\newcommand\nc{\newcommand} \nc\rc{\renewcommand}

\nc\lat\textit
\nc\ie{\lat{i.e.,\ }} \nc\etal{\lat{et al.\ }} \nc\etc{\lat{etc.\ }}
\nc\eg{\lat{e.g,\ }} \nc\insitu{\lat{in situ}} \nc\QED{\lat{Q.E.D.}}
\nc\cf{cf.\ } \nc\wrt{w.r.t.\ }

\nc\re[1]{(\ref{#1})}

\nc\tensor\mathbf
\nc\Tensor\boldsymbol
\nc\ident{\mathbf1}
\nc\zero{\tensor{0}}

\nc\dd{{\rm d}}
\nc\pd{\partial}
\nc\pdt{\pd_t}

\nc\qrho{\varrho}
\nc\qqq{\tensor{q}}
\nc\qqQ{\tensor{Q}}
\nc\qqJ{\tensor{J}}
\nc\qqx{\Tensor{\zeta}}
\nc\qqxi{\tensor{z}}
\nc\qqC{\tensor{C}}
\nc\SigS{\Sigma_{S}}
\nc\qqv{\tensor{v}}

\nc\hht{\hat{t}}
\nc\hx{\hat{x}}
\nc\hy{\hat{y}}
\nc\hdt{\pd_{\hht}}
\nc\hdx{\pd_{\hx}}
\nc\hdy{\pd_{\hy}}
\nc\hdxx{\pd_{\hx \hx}}
\nc\hdyy{\pd_{\hy \hy}}
\nc\hdxy{\pd_{\hx \hy}}

\nc\hq{\hat{q}}
\nc\hQ{\hat{Q}}
\nc\hT{\hat{T}}
\nc\qq[3]{\left( \hq_{#1} \right)^{#2}_{#3}}
\nc\qQ[3]{\left( \hQ_{#1} \right)^{#2}_{#3}}

\nc\dt{\Delta \hht}
\nc\dx{\Delta \hx}
\nc\dy{\Delta \hy}

\nc\half{\nicefrac{1}{2}}

\title{Investigating the whirling heat current density in the Guyer--Krumhansl equation}
\author[1,2,3]{Mátyás Szücs}
\author[1,4]{Carmelo Filippo Munafo}
\author[1,2,3]{Róbert Kovács}

\affil[1]{Department of Energy Engineering, Faculty of Mechanical Engineering, Budapest University of Technology and Economics, Műegyetem rkp. 3., H-1111 Budapest, Hungary}
\affil[2]{Department of Theoretical Physics, HUN-REN Wigner Research Centre for Physics, Institute for Particle and Nuclear Physics, Konkoly Thege Mikl\'{o}s St. 29--33, H-1121 Budapest, Hungary}
\affil[3]{Montavid Thermodynamic Research Group, Budapest, Hungary}
\affil[4]{Department of Mathematical and Computer Sciences, Physical Sciences and Earth Sciences, University of Messina, Viale F. Stagno d’Alcontres 31, 98166, Messina, Italy}

\date{\today}

\begin{document}

\maketitle

\begin{abstract}
Among the numerous heat conduction models, the Guyer--Krumhansl equation has a special role. Besides its various application possibilities in nanotechnology, cryotechnology, and even in case of modeling heterogeneous materials, it poses additional mathematical challenges compared to the Fourier or Cattaneo {(a.k.a. Maxwell--Cattaneo--Vernotte)} equations. Furthermore, the Guyer--Krumhansl equation is the first heat conduction model, which includes the curl of the heat flux density in the evolution equation. In the present paper, we place our focus on the consequences of the existence of such whirling heat current density by solving the two-dimensional Guyer--Krumhansl equation with a space and time-dependent heat pulse boundary condition. The discretization poses further challenges in regard to the boundary condition for which we propose a particular extrapolation method. Furthermore, with the help of the Helmholtz decomposition, we show the analogy with the linearized acoustics of Newtonian fluids, which reveals how the heat flux density plays the role of the velocity field. Our solutions also reveal an unexpected temperature evolution caused by the whirling heat flux density, namely, the temperature can locally be decreased for a short time in a case when the curl of the heat flux density dominates the heat conduction process.
\end{abstract}

\section{Introduction}
Several heat conduction models and approaches have been developed and tested in previous decades, such as the Cattaneo \cite{Cattaneo58}, Guyer--Krumhansl (GK) \cite{GK66}, two-temperature \cite{Sobolev94, Sobolev97, Sobolev16}, and Jeffreys equations \cite{JosPre89}. From an engineering point of view, the GK and Jeffreys equations could be viable alternatives to the Fourier equation \cite{FehKov24}. In some particular situation, also the two-temperature model can be useful, however, with strict restrictions on the composition, time scales and measurements \cite{KovFehSob22}. Both models contain Fourier's law as a particular case, furthermore, they can model second sound (similarly to the Cattaneo equation) and provide a useful, effective description for heterogeneous materials \cite{FehKov24} using a continuum approach. Therefore, the continuum background of GK equation hold broad potential for more and more advanced practical applications, also including heat foam-based heat exchangers and thermal storage technologies \cite{ChenEtal21, NematEtal22, ZhangEtal22}.

The GK equation, however, is a notably more complicated model than the Jeffreys equation, especially in two or three spatial dimensions. 
The reason lies in their background. The Jeffreys equation doubles Fourier's law and does not introduce second-order tensors along with further complications (couplings, isotropic representation, boundary conditions, etc.). On the contrary, the GK equation originates from phonon hydrodynamics \cite{GK66, MulRug98, DreStr93a}, and therefore, it is a special fluid model, thus allowing the curl of the heat flux. This phenomenon is particularly interesting for superfluids \cite{MongEtal18, SykEtal21, JouEtal11}, and also makes us revisit our expectations about how it can appear in the temperature history, even for solids in their effective description. In our work, we place the emphasis on the transient evolution which highlights particular temperature decrease effects occurring locally due to the heat current vorticity in agreement with the findings of \cite{SykEtal23}. This is apparent only in a transient setting, the time evolution of the heat current vorticity induces the local temperature decrease relative to the initial state. Therefore, these observations remain hidden in the work of Beardo et al.~\cite{JordiEtal24} since they applied the GK equation on a stationary problem.

The phonon hydrodynamic description is mainly characteristic of Rational Extended Thermodynamics, constraining the coefficients and possible nonlinearities (state dependence) \cite{MulRug98}. However, it is possible to derive the GK equation in a continuum framework using the internal variable approach. We wish to introduce an additional interesting aspect compared to the work of Fülöp and Ván \cite{VanFul12} in regard to the constitutive properties of the internal variable \cite{BerVan17b}. The continuum background offers more flexibility but keeps the structure of the GK equation. Consequently, as the model becomes free from the limitations of phonon hydrodynamics, the coefficients are merely restricted by the second law of thermodynamics. Thus, the GK equation can be used in the effective description of heterogeneous materials \cite{FehKov24} and still can inherit the coefficients from phonon hydrodynamics. However, we wish to emphasize that we do not want to delve into the general details of two-dimensional materials and the related phonon hydrodynamic properties. In this respect, we want to refer to the recent work of Shang et al.~\cite{ShangEtal20} in which they also conduct a similar investigation as ours, but they slightly modify the GK equation based on particular phonon hydrodynamic assumptions being characteristic at nanoscale for two-dimensional materials. Since phonon hydrodynamics strictly restrict the applicable parameters, the effects of the whirling heat current density remain hidden, however, the flexibility given the present continuum approach allows to find further insights. 

In the present paper, we show the continuum derivation of the GK equation together with its complete isotropic representation -- which, to the best of our knowledge, was first derived by Ván \cite{Van2001} --, highlighting the differences compared to phonon hydrodynamics. Longitudinal and transversal modes are also discussed, highlighting the role of the rotational part of the heat flux field and providing further insights into the structure of the GK equation. Additionally, we solve a two-dimensional problem with space and time-dependent heat pulse boundary condition, using a staggered scheme. Our particular interest is to show the consequences of space-dependent boundaries and how they affect their numerical realization. We evaluate the findings in relation to the curl of the heat flux, the necessary boundary conditions, and how such a complex model can relate to the Fourier equation.


\section{The Guyer--Krumhansl equation within the frame of internal variables}

Considering a rigid, homogeneous, and isotropic heat conductor at rest w.r.t. a given reference frame, the balance of internal energy with neglected mechanical interactions and volumetric heat sources reads as
\begin{align}
    \label{eq:bal-e}
    \qrho \pdt e = - \nabla \cdot \qqq ,
\end{align}
where $ \qrho $, $ e $ and $ \qqq $ are the (mass) density -- which is constant due to the assumption of rigidity --, the (mass) specific internal energy and the heat current density, respectively, $ \pdt $ is the partial time derivative and $ \nabla $ is the nabla operator. {Let $ \Tensor{\varphi} $ denote a physical quantity with arbitrary tensorial order, \ie its components \wrt the corresponding basis built from Cartesian base vectors $ \tensor{e}_j , \ j = 1,2,3 $ are denoted as $ \varphi_{i_{1} \dots i_{N}} , \ i_{1} , \dots , i_{N} = 1,2,3 $.  In accordance with the usual notation and usage of the nabla operator in continuum mechanics, \textit{right} gradient, \textit{right} divergence and \textit{right} curl are defined as
\begin{align}
	\operatorname{grad}_{\rm R} \Tensor{\varphi} &:= \Tensor{\varphi} \otimes \nabla  = \pd_j \varphi_{i_{1} \dots i_{N}} \tensor{e}_{i_1} \otimes \dots \otimes \tensor{e}_{i_N} \otimes \tensor{e}_{j}, \\
	\operatorname{div}_{\rm R} \Tensor{\varphi} &:= \Tensor{\varphi} \cdot \nabla = \pd_j \varphi_{i_{1} \dots i_{N - 1} j} \tensor{e}_{i_1} \otimes \dots \otimes \tensor{e}_{i_{N-1}} , \\
	\operatorname{curl}_{\rm R} \Tensor{\varphi} &:= \Tensor{\varphi} \times \nabla  = \epsilon_{i_{N} j k} \pd_j \varphi_{i_{1} \dots i_{N}} \tensor{e}_{i_1} \otimes \dots \otimes \tensor{e}_{i_{N -1}} \otimes \tensor{e}_{k} ,
\end{align}
where Einstein summation is applied and $ \epsilon_{ijk} $ is the Levi-Civita permutation symbol. Especially, for an arbitrary scalar field $ a $
\begin{align}
	a \otimes \nabla &= \nabla \otimes a =: \nabla a
\end{align}
and, divergence and rotation is not interpreted. For an arbitrary vector field $ \tensor{v} $
\begin{align}
	\tensor{v} \otimes \nabla &= \left( \nabla \otimes \tensor{v} \right)^{\rm T} , \\
	\tensor{v} \cdot \nabla &= \nabla \cdot \tensor{v} , \\
	\tensor{v} \times \nabla &= - \nabla \times \tensor{v} ,
\end{align}
where $ ^{\rm T} $ denotes the transpose of a second order tensor.}

Assuming that the thermodynamical state space is spanned by the specific internal energy and an internal variable denoted by $ \qqx $, then the Gibbs relation for the specific entropy function $ s $ reads as
\begin{align}
\label{eq:Gibbs}
    \dd s = \frac{1}{T} \dd e - \qqxi \cdot \dd \qqx ,
\end{align}
\ie the partial derivatives of entropy w.r.t. variables $ e $ and $ \qqx $ are
\begin{align}
    \frac{\pd s}{\pd e} &= \frac{1}{T} , &
    \frac{\pd s}{\pd \qqx} &= - \qqxi ,
\end{align}
where $ T $ denotes the temperature and $ \qqxi $ is the entropy conjugate of the internal variable.
Let us assume that entropy current density $ \qqJ $ is also generalized through a second order tensor, called Nyíri multiplier \cite{Nyiri1991} and denoted by $ \qqC $, \ie
\begin{align}
\label{eq:Nyiri}
    \qqJ = \left( \frac{1}{T} \ident + \qqC \right) \qqq ,
\end{align}
where $ \ident $ is the identity tensor. Therefore, entropy production rate density $ \SigS $ is calculated through the balance of entropy
\begin{align}
    \nonumber
    0 \le \SigS &= \qrho \pdt s + \nabla \cdot \qqJ \stackrel{\re{eq:Gibbs}-\re{eq:Nyiri}}{=} \frac{1}{T} \qrho \pdt e - \qqxi \cdot \qrho \pdt \qqx + \left[ \left( \frac{1}{T} \ident + \qqC \right) \cdot \nabla \right] \cdot \qqq + \left( \frac{1}{T} \ident + \qqC \right) : \left( \qqq \otimes \nabla \right) \\
    \nonumber
    &\stackrel{\re{eq:bal-e}}{=} - \frac{1}{T} \nabla \cdot \qqq - \qqxi \cdot \qrho \pdt \qqx + \left[ \left( \frac{1}{T} \ident + \qqC \right) \cdot \nabla \right] \cdot \qqq + \frac{1}{T} \nabla \cdot \qqq + \qqC : \left( \qqq \otimes \nabla \right) \\
    \label{eq:ent-prod}
    &= - \qqxi \cdot \qrho \pdt \qqx + \left[ \left( \frac{1}{T} \ident + \qqC \right) \cdot \nabla \right] \cdot \qqq + \qqC : \left( \qqq \otimes \nabla \right) .
\end{align}
According to classical irreversible thermodynamics \cite{deGroot1963,Gyarmati1970}, entropy production rate density is a quadratic expression of thermodynamical fluxes and forces, among which functional relations are prescibed, \ie fluxes depend on the forces. However, to differentiate between fluxes and forces in \re{eq:ent-prod} is complicated, these correspondences are rather formal. Nevertheless, the proven Onsagerian method works, we choose and group the quantities according to Table \ref{tab:1}.
\begin{table}[h!]
\centering
\begin{tabular}{l || c | c | c}
To be determined (`fluxes') & $ \qrho \pdt \qqx $ & $ \qqq $ & $ \qqC $ \\
\hline
In relation with (`forces') & $ \qqxi $ & $ \left( \frac{1}{T} \ident + \qqC \right) \cdot \nabla $ & $ \qqq \otimes \nabla $
\end{tabular}
\caption{Thermodynamic `fluxes' and `forces'.}
\label{tab:1}
\end{table}
 
Assuming isotropic material, due to Curie's principle the different tensorial orders and characters do not couple, hence positive semi-definiteness of \re{eq:ent-prod} is ensured via the linear Onsagerian equations
\begin{align}
    \label{eq:Ons-1}
    \qrho \pdt \qqx &= - l_{11} \qqxi + l_{12} \left( \frac{1}{T} \ident + \qqC \right) \cdot \nabla , \\
    \label{eq:Ons-2}
    \qqq &= - l_{21} \qqxi + l_{22} \left( \frac{1}{T} \ident + \qqC \right) \cdot \nabla, \\
    \label{eq:Ons-3}
    \qqC &= \mathbb{L} : \left( \qqq \otimes \nabla \right)
\end{align}
with the criteria on the scalar coefficients
\begin{align}
    l_{11} & \ge 0 , & l_{22} & \ge 0 , & l_{11} l_{22} - l_{12} l_{21} & \ge 0 
\end{align}
and the fourth order constitutive tensor $ \mathbb{L} $ with appropriate conditions. Since isotropy of the material is assumed \re{eq:Ons-3} can be given in the isotropic decomposition as
\begin{align}
    \label{eq:Ons-3-iso}
    \qqC &= \frac{L^{\rm s} - L^{\rm d}}{3} \left( \nabla \cdot \qqq \right) \ident + \frac{L^{\rm d}}{2} \left( \qqq \otimes \nabla + \nabla \otimes \qqq \right)  + \frac{L^{\rm A}}{2} \Tensor{\epsilon} : \left( \qqq \times \nabla \right) ,
\end{align}
with the Levi--Civita tensor $ \Tensor{\epsilon} $ and the scalar coefficients
\begin{align}
    L^{\rm s} & \ge 0 , & L^{\rm d} & \ge 0 , & L^{\rm A} & \ge 0 ,
\end{align}
which map the spherical, symmetric traceless (\ie deviatoric), and antisymmetric parts of the gradient of heat current density, respectively. {In general, coefficients $ l_{11} $, $ l_{12} $, $ l_{21} $, $ l_{22} $, $ L^{\rm s} $, $ L^{\rm d} $ and $ L^{\rm A} $ are state-dependent parameters, however, for simplicity, we treat them as constants.} Equations \re{eq:Ons-1} and \re{eq:Ons-2} can be reformulated as
\begin{align}
    \label{eq:Ons-11}
    \qrho \pdt \qqx &= - l_{11} \qqxi + l^{\rm S} \left( \frac{1}{T} \ident + \qqC \right) \cdot \nabla + l^{\rm A} \left( \frac{1}{T} \ident + \qqC \right) \cdot \nabla , \\
    \label{eq:Ons-22}
    \qqq &= - l^{\rm S} \qqxi + l^{\rm A} \qqxi + l_{22} \left( \frac{1}{T} \ident + \qqC \right) \cdot \nabla
\end{align}
with $ l^{\rm S} = \frac{1}{2} \left( l_{12} + l_{21} \right) $ and $ l^{\rm A} = \frac{1}{2} \left( l_{12} - l_{21} \right) $, then entropy production rate density can be given as
\begin{align}
    \nonumber
    0 \le \SigS =
    \begin{pmatrix} - \qqxi & \frac{1}{T} \ident + \qqC \end{pmatrix} \cdot
    \begin{pmatrix}
        l_{11} & l^{\rm S} \\
        l^{\rm S} & l_{22}
    \end{pmatrix}
    \begin{pmatrix} - \qqxi \\ \frac{1}{T} \ident + \qqC \end{pmatrix} + \qqC : \left( \qqq \otimes \nabla \right) ,
\end{align}
therefore, the terms in \re{eq:Ons-11} and \re{eq:Ons-22} with coefficients $ l^{\rm A} $ do not increase entropy. Let us now assume that $ l^{\rm A} = -1 $, $ l^{\rm S} = 0 $ and $ l_{22} = 0 $, then one obtains for \re{eq:Ons-11} and \re{eq:Ons-22}
\begin{align}
    \label{eq:Ons-111}
    \qrho \pdt \qqx &= - \left( \frac{1}{T} \ident + \qqC \right) \cdot \nabla - l_{11} \qqxi , \\
    \label{eq:Ons-222}
    \qqq &= - \qqxi .
\end{align}
According to the latter equation, the entropy-conjugated internal variable $ - \qqxi $ is identified with the heat current density. Furthermore, assuming the linear equation of state 
\begin{align}
    \label{eq:const-eq-xi}
    \qqxi = m \qqx
\end{align}
{with a constant $ m $ -- which is positive to ensure concavity of specific entropy --}, the internal variable can be eliminated, hence \re{eq:Ons-111} and \re{eq:Ons-222} reduces to the single equation
\begin{align}
    \label{eq:almost-GK}
    - \frac{\qrho}{m} \pdt \qqq &= - \left( \frac{1}{T} \ident + \qqC \right) \cdot \nabla + l_{11} \qqq .
\end{align}
Finally, replacing \re{eq:Ons-3-iso} into \re{eq:almost-GK}, we obtain the Guyer--Krumhansl equation
\begin{align}
    \label{eq:GK}
    \tau \pdt \qqq + \qqq = - \lambda \nabla T + \eta_1 \Delta \qqq + \eta_2 \nabla \left( \nabla \cdot \qqq \right)
\end{align}
with the coefficients defined as
\begin{align} 
\label{eq:coeff}
    \lambda &:= \frac{ 1 }{ l_{11} T^2 } \ge 0 , &
    \tau &:= \frac{\qrho}{m l_{11}} \ge 0 , &
    \eta_1 &:= \frac{ L^{\rm d} + L^{\rm A} }{ 2 l_{11} } \ge 0 , &
    \eta_2 &:= \frac{ 2 L^{\rm s} + L^{\rm d} - 3 L^{\rm A} }{ 6 l_{11} } .
\end{align}
The balance of internal energy \re{eq:bal-e}, the Guyer--Krumhansl equation together with the (thermostatic caloric) equation of state
\begin{align}
    \label{eq:eos}
    e = c T ,
\end{align}
where $ c $ is the specific heat capacity of the material, forming a closed system of equations, which, with appropriate initial and boundary conditions, can be solved.

We must observe that $\eta_1$ and $\eta_2$ are linearly independent coefficients and are merely restricted by the second law of thermodynamics. Therefore, they are positive semidefinite and free from the restrictions given by the phonon hydrodynamic background, \ie $\eta_1$ is not identical to the square of the mean free path, and the ratio of $\eta_2 $ to $ \eta_1 $ is not necessarily 2 either. The present continuum background provides a notably more flexible adjustment for the coefficients, either by means of an experiment such as \cite{FehKov24} or inheriting the phonon hydrodynamic approach \cite{MulRug98, DreStr93a}. 

Furthermore, we wish to call attention to the coefficients \eqref{eq:coeff} in which relations their functional relationships become apparent. It is clear that if $\lambda=\lambda(T)$ holds (\eg a linear or exponential one), then $l_{11}=l_{11}(T)$ can be immediately given and that $T$-dependence is inherited in all the other coefficients. The different parameters can optionally adjust the necessary $T$-dependence. Additionally, if such nonlinearities are required, then equations \re{eq:Ons-3-iso}, \re{eq:Ons-11} and \re{eq:Ons-22} [or instead of these latter ones \re{eq:Ons-111} and \re{eq:Ons-222}] must be the starting point in order to take into account the additional contributions correctly. For instance, due to an assumed temperature dependence in the coefficients $ L^{\rm s} $, $ L^{\rm d} $ and $ L^{\rm A} $ a direct multiplicative coupling between temperature gradient and gradient of heat current density emerges.

Comparing our derivation to the work of Fülöp and Ván \cite{VanFul12}, we have introduced two notable modifications here.

First, we distinguished between the internal variable and its entropy conjugate, revealing the need for an additional equation of state. {In other words, the particular form of specific entropy
\begin{align}
	\label{eq:s-KovVan}
    s(e,\qqx) = \hat s(e) - \frac{m}{2} \qqx \cdot \qqx
\end{align}
with $ m > 0 $ constant immediately implicitly imposes equation \eqref{eq:const-eq-xi}, hence, in this linear case with constant coefficients the two derivations are equivalent. When the equation of state \re{eq:const-eq-xi} is assumed to be linear, then the internal variable itself can be identified with the heat current density, as done, for example, in \cite{KovVan2015}. Futhermore, distinction between the internal variable and its conjugate enables the compatibility with the GENERIC (General Equation for the Non-Equilibrium Reversible--Irreversible Coupling) framework \cite{SzucsEtal2022}. Finally we note that if equation of state \re{eq:const-eq-xi} is beyond a linear relationship, then elimination of the internal variable and its conjugate can be complicated or impossible, therefore, in this case deeper knowledge (\eg microscopic or mesoscopic interpretations) about these variables are required.}

Second, we interpret the current multiplier as a separate, relaxed state variable in accordance with \cite{SzucsEtal2022}, hence a direct interpretation of Müller's $ \tensor{K} $-vector {as $ \tensor{K} = \qqC \qqq $} is obtained \cite{Muller1968,Muller1971}.

\section{Longitudinal and transversal heat propagation}

Via the vector Laplacian identity
\begin{align}
    \Delta \qqq = \nabla \left( \nabla \cdot \qqq \right) - \nabla \times \nabla \times \qqq 
\end{align}
the Guyer--Krumhansl constitutive equation \re{eq:GK} can also be given as
\begin{align}
    \label{eq:GK-curl}
    \tau \pdt \qqq + \qqq = - \lambda \nabla T + \left( \eta_1 + \eta_2 \right) \nabla \left( \nabla \cdot \qqq \right) - \eta_1 \nabla \times \nabla \times \qqq ,
\end{align}
therefore, theoretically, in a GK-type heat conductor transversal heat propagation can also be observed. {Via the Helmholtz decomposition, the heat current density can be uniquely -- up to a space-independent but arbitrary time-dependent vector function -- given as a sum of irrotational (\ie curl-free, $\qqq^*$) and solenoidal (\ie divergence-free, $\qqq^\circ$) vector fields, \ie}
\begin{align}
    \qqq &= \qqq^* + \qqq^\circ , &
    \text{with} &&
    \nabla \cdot \qqq &= \nabla \cdot \qqq^* &
    \text{and} &&
    \nabla \times \qqq &= \nabla \times \qqq^\circ .
\end{align}
Curl-free and divergence-free components of the vector field are usually referred to as longitudinal and transversal components, respectively. Since $ \nabla T $ is curl-free, the governing equations [together with \re{eq:eos}] of a GK-type heat conductor can be reformulated as
\begin{align}
     \label{eq:gkT}
    \qrho c \pdt T &= - \nabla \cdot \qqq^* , \\
     \label{eq:gkq}
    \tau \pdt \qqq^* + \qqq^* &= - \lambda \nabla T + \left( \eta_1 + \eta_2 \right) \Delta \qqq^* , \\
    \tau \pdt \qqq^\circ + \qqq^\circ &= \eta_1 \Delta \qqq^\circ , \label{eq:gk0}
\end{align}
which decomposition reveals the time evolutions of the longitudinal and transversal components of heat current density. Here, the evolution of $\qqq^\circ$ is decoupled from evolutions of $\qqq^*$ and $T$, thus $\qqq^\circ$ can only be introduced through a spatially-dependent boundary condition or by a particular initial condition. A non-homogeneous temperature field alone can not induce non-zero $ \qqq^\circ $. Since boundary conditions on the heat current density defines $\qqq$ itself, it is inevitable to introduce disturbances into both parts of $\qqq$, but unique separation of these on the boundary can be complicated.

Later on, we refer on $ \eta_1 + \eta_2 $ as the longitudinal GK coefficient and on $ \eta_1 $ as the transversal GK coefficient. Based on our previous 1 spatial dimensional studies of the GK equation \cite{FehEtal21}, the longitudinal GK coefficient (denoted usually in 1 spatial dimension with $ \kappa^2 $) has to be positive semi-definite, hence the seemingly indefinite $ \eta_2 $ parameter in \re{eq:coeff} is constraint by
\begin{align}
	\eta_2 \ge - \eta_1 .
\end{align}

{An additional interesting property is related to the Fourier resonance condition, \ie the temperature history given by the GK equation is identical with the Fourier's one. Namely, replacing the gradient of \re{eq:gkT} into the partial time derivative of \re{eq:gkq} and taking the advantage of commutation of $ \nabla $ and $ \pdt $, one obtains
\begin{align}
	\tau \pdt \left( \pdt \qqq^* - \frac{\eta_1 + \eta_2}{\tau} \Delta \qqq^* \right) + \left( \pdt \qqq^* - \frac{\lambda}{\qrho c} \Delta \qqq^* \right) &= 0 ,
\end{align}
which is the sum of a Fourier heat conduction equation and the partial time derivative of a slightly modified Fourier heat conduction equation. If the additional time scale becomes identical given by the thermal diffusivity, \ie
\begin{align}
	\label{eq:Fres}
 \frac{\eta_1 + \eta_2}{\tau} = \frac{\lambda}{\qrho c} ,
\end{align}
Fourier resonance occurs. The resonance condition \re{eq:Fres} is the same as obtained in the one-dimensional case \cite{FehEtal21}, however, in the three-dimensional setting, Fourier resonance appears only in the longitudinal direction, the transversal contribution may distort this behaviour. When
\begin{align}
	\frac{\left( \eta_1 + \eta_2 \right) / \tau}{\lambda / (\qrho c)} > 1,
\end{align}
then over-diffusive solutions are obtained, while in the under-damped opposite case attenuated wave-like propagation of the temperature field is observable. From an engineering point of view, Fourier resonance also seems to be a natural requirement, since the GK equation also consists of the Fourier equation as a particular case, and thus it is highly advantageous in practice if the GK equation can reproduce the simpler Fourier solution by the particular alignment of coefficients, without any modifications on the model.}

Let us refer here on the governing equations of linearized acoustics of Newtonian fluids, which reads -- via applying Helmholtz decomposition on the velocity field $ \qqv $ -- 
\begin{align}
    \label{eq:ac-1}
    \pdt \rho &= - \bar \rho \nabla \cdot \qqv^* , \\
    \label{eq:ac-2}
    \bar \rho \pdt \qqv^* &= - \bar{a}_s^2 \nabla \rho + \left( \bar{\eta}_{\rm Vol} + \frac{4}{3} \bar{\eta}_{\rm Sh} \right) \Delta \qqv^* , \\
    \label{eq:ac-3}
    \bar \rho \pdt \qqv^\circ &= \bar{\eta}_{\rm Sh} \Delta \qqv^\circ
\end{align}
with density difference $ \rho $ measured from density $ \bar \rho $ of the unperturbed state, isentropic speed of sound $ \bar{a}_s $, volume and shear viscosities $ \bar{\eta}_{\rm Vol} $ and $ \bar{\eta}_{\rm Sh} $, respectively (\cf equations (3.100) and (3.109) in \cite{Rossing2007}). Via the curl of velocity the vorticity $ \Tensor{\omega} = \nabla \times \qqv = \nabla \times \qqv^\circ $ is defined, hence \re{eq:ac-3} can also be written on $ \Tensor{\omega} $ instead of $ \qqv^\circ $. Therefore, Helmholtz decomposition highlights that in linear approximation of acoustics the evolution of $ \Tensor{\omega} $ is decoupled from $ \rho $ and $ \qqv^* $, hence resulting that sound is a transversal wave [\cf \re{eq:ac-1} and \re{eq:ac-2}] and vorticity can not be introduced through the acoustic fields $ \rho $ and $ \qqv^* $ (nor through the pressure field). We are dealing with something very similar in case of GK equations \re{eq:gkT}, \re{eq:gkq} and \re{eq:gk0}.

\section{A staggered grid finite difference method demonstrated on the heat pulse experiment in three spatial dimensions}

{We aim to numerically model the heat pulse experiment in which a single short pulse thermally excites the sample. We take into account both the spatial and time dependence of the pulse, serving as an outstanding example to demonstrate the role of boundary conditions due to the appearance of in-plane derivatives. }

In order to ease the discretization of the second-order derivative of $\qqq$, it is advantageous to introduce the gradient of heat current density as an auxiliary variable
\begin{align}
    \qqQ := \qqq \otimes \nabla ,
\end{align}
hence, a system of first-order equations has to be solved. Additionally, that $\qqQ$ feels natural to introduce in the case of the GK equation as it strongly resembles the current multiplier $\qqC$. After that reformulation, we obtain
\begin{align}
    \label{eq:GK2}
    \tau \pdt \qqq + \qqq = - \lambda \nabla T + \eta_1 \qqQ \cdot \nabla + \eta_2 \nabla \operatorname{tr} \qqQ
\end{align}
for which it is crucial to correctly discretize $\qqQ$ on the boundary. 

We wish to solve the GK equation in Cartesian coordinate system for a rectangular domain with the size of $ X \times Y \times Z $. We assume that the heat pulse excites the entire $Z$ direction uniformly, therefore we can reduce the problem to two-dimensional in $X$ and $Y$. Furthermore, we consider symmetry at $y=0$, so that $ - \frac{Y}{2} \le y \le \frac{Y}{2} $, and thus we deal only half of the rectangle. Consequently, the boundary conditions are 
\begin{align}
    q_x ( t , x = 0 , y ) &=
    \begin{cases}
         Q_{ \rm P,Z } \frac{1}{\tau_{ \rm P }^{} Y_{ \rm P }^{} } \left[ 1 - \cos \left( 2 \pi \frac{ t }{ \tau_{ \rm P }^{} } \right) \right] \left[ 1 + \cos \left( 2 \pi \frac{y}{Y_{ \rm P }^{}}\right) \right] & \text{if\ } 0 \le t \le \tau_{ \rm P }^{} \text{\ and\ } 0 \le y \le \frac{Y_{ \rm P }^{}}{2} \le \frac{Y}{2}  \\
        0 & \text{otherwise}
    \end{cases} , \\
    q_x ( t , x = X , y ) &= 0 , \\
    q_y ( t , x , y = 0 ) &= 0 , \\
    q_y ( t , x , y = + \frac{Y}{2} ) &= 0 .
\end{align}
where $ Q_{\rm P,Z} $ is the $ Z $-length specific amount of heat introduced during the heat pulse, measured in $ \rm \frac{J}{m} $.
The initial condition describes equilibrium with homogeneous temperature distribution $T(t=0,x,y)=T_0$, hence $\qqq(t=0,x,y)=\mathbf 0$, and $\qqQ(t,x,y)=\mathbf 0$.

We wish to transform the GK equation to a non-dimensional one using the following characteristic scales. We choose $X$ for the length scale and $ \frac{X^2}{\alpha} $ for the time scale using the thermal diffusivity $ \alpha := \frac{\lambda}{\qrho c} $ (this leads to the usual Fourier number).
Therefore, we obtain the following non-dimensional variables
\begin{align}
    \hht &:= \frac{t}{\frac{X^2}{\alpha}} , &
    \hx &:= \frac{x}{X} , &
    \hy &:= \frac{y}{X},
\end{align}
and derivatives
\begin{align}
    \hdt &= \frac{X^2}{\alpha} \pdt , &
    \hdx &= X \pd_x , &
    \hdy &= X \pd_y .
\end{align}
The non-dimensional fields read
\begin{align}
    \hq_{i} := \frac{q_{i}}{\frac{\alpha Q_{\rm P,Z}}{X^3 R_Y}} , \quad \hQ_{ij} = \frac{Q_{ij}}{\frac{\alpha Q_{\rm P,Z}}{X^2 R_Y}}, \quad \hT := \frac{T - T_0}{T_{\rm max} - T_0} = \frac{\qrho c X^2 R_Y \left( T - T_0 \right)}{Q_{\rm P,Z}}
\end{align}
in which $i, j = x, y$, $ R_Y = \frac{Y}{X} $ and $ T_{\rm max} $ is calculated via integrating \re{eq:bal-e} [together with \re{eq:eos}] on the whole sample volume in time from the initial homogeneous temperature state $ T_0 $ to the final homogeneous temperature state $ T_{\rm max} $.

%
Let us summarize the complete system of non-dimensional equations, 
\begin{align}
    \label{eq:nondim-1}
    \hdt \hT &= - \left( \hdx \hq_x + \hdy \hq_y \right) , \\
    \hat{\tau} \hdt \hq_x + \hq_x &= - \hdx \hT + \left( \hat{\eta}_1 + \hat{\eta}_2 \right) \hdx \hQ_{xx} + \hat{\eta}_1 \hdy \hQ_{xy} + \hat{\eta}_2 \hdx \hQ_{yy} , \\
    \hat{\tau} \hdt \hq_y + \hq_y &= - \hdy \hT + \left( \hat{\eta}_1 + \hat{\eta}_2 \right) \hdy \hQ_{yy} + \hat{\eta}_1 \hdx \hQ_{yx} + \hat{\eta}_2 \hdy \hQ_{xx} , \\
    \label{eq:nondim-2}
    \hQ_{xx} &= \hdx \hq_x , \\
    \hQ_{xy} &= \hdy \hq_x , \\
    \hQ_{yx} &= \hdx \hq_y , \\
    \label{eq:nondim-n}
    \hQ_{yy} &= \hdy \hq_y
\end{align}
where
\begin{align}
    \hat{\tau} &= \frac{\tau}{\frac{X^2}{\alpha}} , &
    \hat{\eta}_1 &= \frac{\eta_1}{X^2} , &
    \hat{\eta}_2 &= \frac{\eta_2}{X^2},
\end{align}
and the non-dimensional heat pulse boundary conditions reads
\begin{align}
    \hq_x ( \hht , \hx = 0 , \hy ) &=
    \begin{cases}
         \frac{R_Y}{\hat{\tau}_{ \rm P }^{} R_{ Y,\rm P }^{} } \left[ 1 - \cos \left( 2 \pi \frac{ \hht }{ \hat{\tau}_{ \rm P }^{} } \right) \right] \left[ 1 + \cos \left( 2 \pi \frac{\hy}{R_{ Y,\rm P }^{}}\right) \right] & \text{if\ } 0 \le \hht \le \hat{\tau}_{ \rm P }^{} \text{\ and\ } 0 \le \hy \le R_{Y , \rm P }^{} \le R_Y  \\
        0 & \text{otherwise}
    \end{cases} ,
\end{align}
{where $ R_{ Y,\rm P } = \frac{Y_{ \rm P }}{X} $ stands.}

\begin{figure}[H]
    \centering
     \includegraphics[width=1\textwidth]{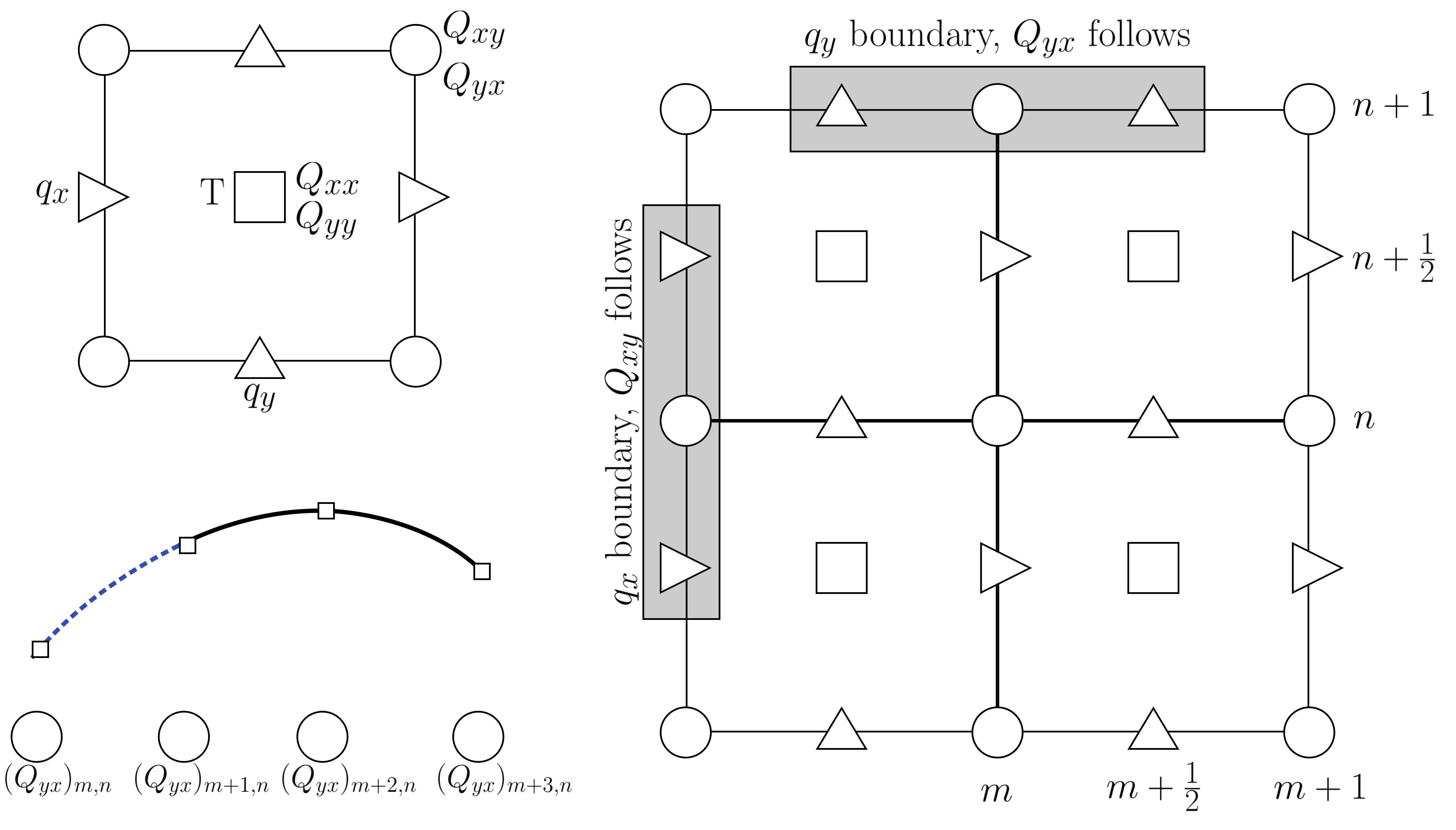}
    \caption{Discretization of the 2D GK equation, showing the staggered grid, and how $\qqQ$ points are allocated and extrapolated from the bulk. Additionally, the lower left figure demonstrated the extrapolation we use to determine the unknown off-diagonal elements.}
    \label{fig:discretisation}
\end{figure}

Spatial discretization is realized via a staggered scheme \cite{FulEtal20}, which is depicted in Figure \ref{fig:discretisation}, and the structure of governing equations \re{eq:nondim-1}--\re{eq:nondim-n} restricts how each field can be represented on the discrete lattice {with directional equidistant grid points with distance $ \dx $ and $ \dy $, hence $ \hat x_m = m \dx, \ m = 1 , \dots , M $, $ \hat y_n = n \dy , \ n = 1 , \dots , N $, where $ M = \frac{1}{\dx} $ and $ N = \frac{R_Y}{2 \dy} $. The investigated time interval is also discretized through equidistance time steps $ \dt $, \ie $ \hat t^j = j \dt , \ j = 1 , \dots , J $. Therefore, the approximated value of a function $ f $ in the discrete time and space coordinates $ \left( \hat t^j , \hat x_m, \hat y_n \right) $ is denoted by $ f^j_{m,n} $.} Temperature, as a state variable characterizing homogeneously one discrete cell, is placed in the middle of the cell, while heat current density characterizing fluxes through the boundaries of the cell, therefore, the corresponding normal components of the heat current density are placed on the boundaries of the cell in line with the discrete temperature values. Discretization of $\qqQ$ follows directly from the discrete values of heat current density and equations \re{eq:nondim-2}--\re{eq:nondim-n}, consequently, its diagonal elements are also in the middle of the cell, but its off-diagonal elements are placed in the corners of the cell. Since we apply $\qqq$-boundaries, only complete cells are used to discretize the entire spatial domain. 

Furthermore, let us note that according to Figure \ref{fig:discretisation}, one needs to prescribe $Q_{xy}$ and $Q_{yx}$ on the boundaries, in accordance with the $\qqq$-boundaries. On each side, one of these off-diagonal quantities can be determined analytically and represented on the discrete lattice. For instance, for a given $q_x(t,x=0,y)$, $Q_{xy}$ can be determined immediately, however, $Q_{yx}$ must be extrapolated from the bulk nodes. This procedure holds for all four boundaries. For the extrapolation, we use quadratic Lagrange polynomials in order to preserve the sign given by three bulk points next to each boundary. This extrapolation is schematically demonstrated in Figure \ref{fig:discretisation} for one setting in which the $m^{th}$ value of $Q_{yx}$ is calculated based on the bulk point of $m+1$, $m+2$, and $m+3$. We want to emphasize that any direct definition of the $\qqQ$ on the boundaries can significantly distort the physical content of the solution, and most probably, a different problem is solved than expected in that case. With our method utilizing Lagrange polynomials, we can avoid the definition of any additional, unnecessary boundary conditions or introduction of virtual nodes.

For the time derivatives, we choose the simplest forward time stepping method, the explicit Euler, which is eligible for our aim to investigate the solutions of the two-dimensional GK equation. Summarizing the numerical scheme built in accordance of Figure \ref{fig:discretisation}:

\begin{align}
    \hT_{m+\half , n+\half}^{j+1} &= T_{m+\half , n+\half}^{j} + \dt \left( \frac{\qq{x}{j}{m,n+\half} - \qq{x}{j}{m+1,n+\half}}{\dx} + \frac{\qq{y}{j}{m+\half,n} - \qq{y}{j}{m+\half,n+1}}{\dy} \right) , \\
    \nonumber
    \qq{x}{j+1}{m , n+\half} &= \left( 1 - \frac{\dt}{\hat{\tau}} \right) \qq{x}{j}{m , n+\half} + \frac{\dt}{\hat{\tau}} \Bigg[ \frac{\hT^{j}_{m-\half , n+\half} - \hT^{j}_{m+\half , n+\half}}{\dx} \mathop{+} \\
    \nonumber
    & \hskip 5ex + \left( \hat{\eta}_1 + \hat{\eta}_2 \right) \frac{\qQ{xx}{j}{m+\half , n+\half} - \qQ{xx}{j}{m-\half , n+\half}}{\dx} + \hat{\eta}_1 \frac{\qQ{xy}{j}{m , n+1} - \qQ{xy}{j}{m , n}}{\dy} \mathop{+} \\
    & \hskip 5ex + \hat{\eta}_2  \frac{\qQ{yy}{j}{m+\half , n+\half} - \qQ{yy}{j}{m-\half , n+\half}}{\dx} \Bigg] , \\
    \nonumber
    \qq{y}{j+1}{m+\half , n} &= \left( 1 - \frac{\dt}{\hat{\tau}} \right) \qq{y}{j}{m+\half , n} + \frac{\dt}{\hat{\tau}} \Bigg[ \frac{\hT^{j}_{m+\half , n-\half} - \hT^{j}_{m+\half , n+\half}}{\dx} \mathop{+} \\
    \nonumber
    & \hskip 5ex + \left( \hat{\eta}_1 + \hat{\eta}_2 \right) \frac{\qQ{yy}{j}{m+\half , n+\half} - \qQ{yy}{j}{m+\half , n-\half}}{\dy} + \hat{\eta}_1 \frac{\qQ{yx}{j}{m+1 , n} - \qQ{xy}{j}{m , n}}{\dx} \mathop{+} \\
    & \hskip 5ex + \hat{\eta}_2  \frac{\qQ{xx}{j}{m+\half , n+\half} - \qQ{xx}{j}{m+\half , n-\half}}{\dy} \Bigg] , \\
    \qQ{xx}{j}{m+\half , n+\half} &= \frac{\qq{x}{j}{m+1 , n+\half} - \qq{x}{j}{m , n+\half}}{\dx} , \\
    \qQ{xy}{j}{m , n} &= \frac{\qq{x}{j}{m , n+\half} - \qq{x}{j}{m , n-\half}}{\dy} , \\
    \qQ{yx}{j}{m , n} &= \frac{\qq{y}{j}{m+\half , n} - \qq{y}{j}{m-\half , n}}{\dx} , \\
    \qQ{yy}{j}{m+\half , n+\half} &= \frac{\qq{y}{j}{m+\half , n+1} - \qq{y}{j}{m+\half , n}}{\dy} .
\end{align}

\section{Numerical results}






\subsection{2D Fourier solutions}
\begin{figure}[H]
    \centering
     \includegraphics[width=1\textwidth]{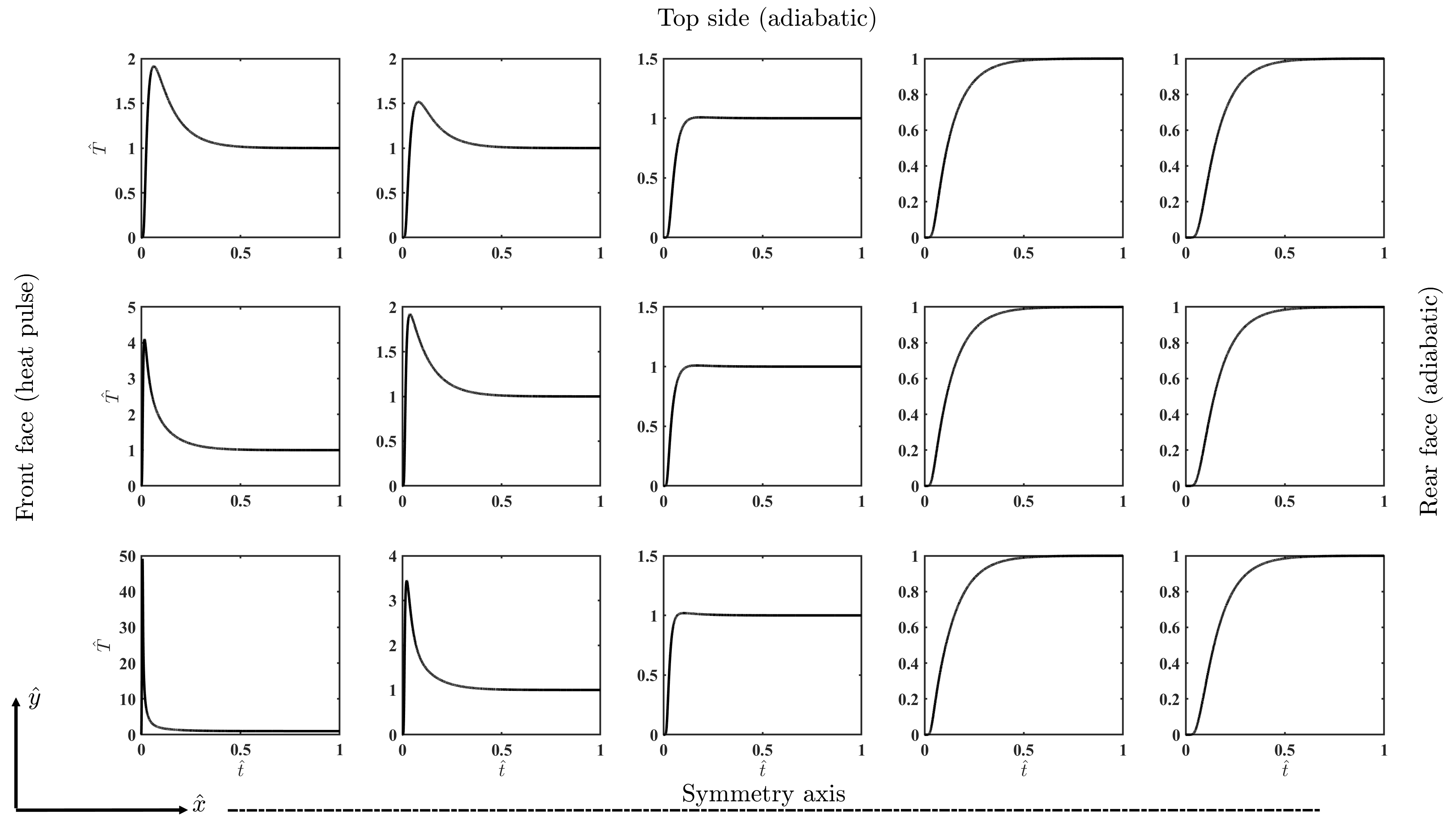}
    \caption{2D Fourier solution in space and time, also indicating the corresponding boundary conditions.}
    \label{fig:F1}
\end{figure}

\begin{figure}[H]
    \centering
     \includegraphics[width=1\textwidth]{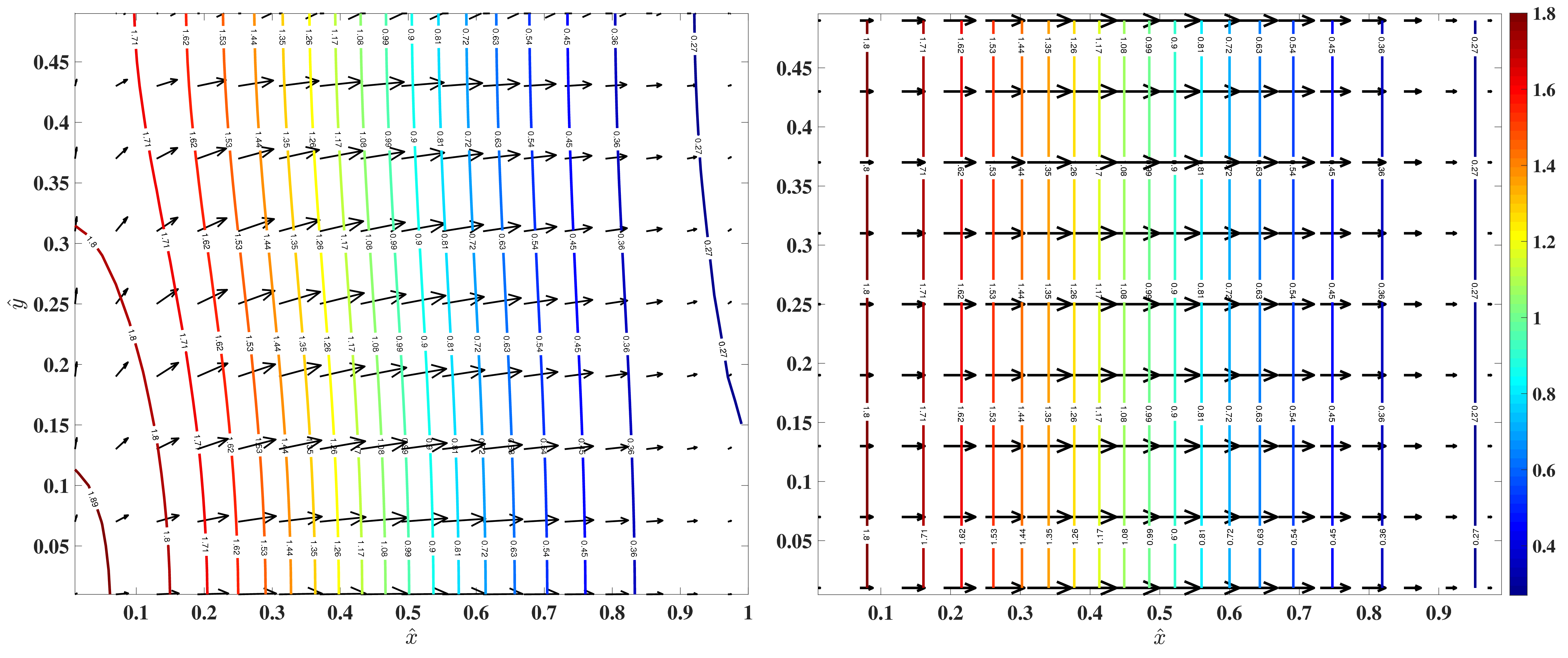}
    \caption{2D vector plot of the heat flux field for the Fourier equation at $\hat t=0.1$, using the same color bar for both figures. Left: spatially inhomogeneous heat pulse boundary condition. Right: spatially homogeneous heat pulse boundary condition.}
    \label{fig:F2&3}
\end{figure}

Here, let us start with the Fourier heat equation and present the plots we use to visualize the temperature and heat flux density history in time and space. We wish to emphasize that we have a different code for the Fourier equation in which we solve only the Fourier equation and not the simplification of the GK equation, hence there are no difficulties with the boundary conditions in this case, \ie no $\qqQ$ is used. Figure \ref{fig:F1} shows the temperature histories at specified spatial points. The first column shows the front face, and each column increases the spatial step by $1/4$, and the last one is related to the rear face of the sample. Additionally, the first row shows the top side, and the last one presents the temperature history on the symmetry axis. We use the same plot concept for the GK equation as well. Furthermore, it is insightful to check the vector plot for the heat flux field in which we can observe the effect of space-dependent excitation and compare it immediately with the corresponding temperature distribution, see Figure \ref{fig:F2&3}. Also, Figure \ref{fig:F2&3} shows that it is straightforward to reproduce the well-known 1D solutions by applying a spatially homogeneous boundary condition. The curl of the heat flux field is identically zero. {For all the subsequent calculations, we fix $R_{ Y,\rm P }=0.4 $ with fixed spatial resolution ($\Delta \hat x = \Delta \hat y=0.02$), and $\hat{\tau}_{ \rm P } = 0.01$.}

\subsection{2D GK solutions}
\subsubsection*{Demonstrating the Fourier resonance} 
\begin{figure}[H]
    \centering
     \includegraphics[width=1\textwidth]{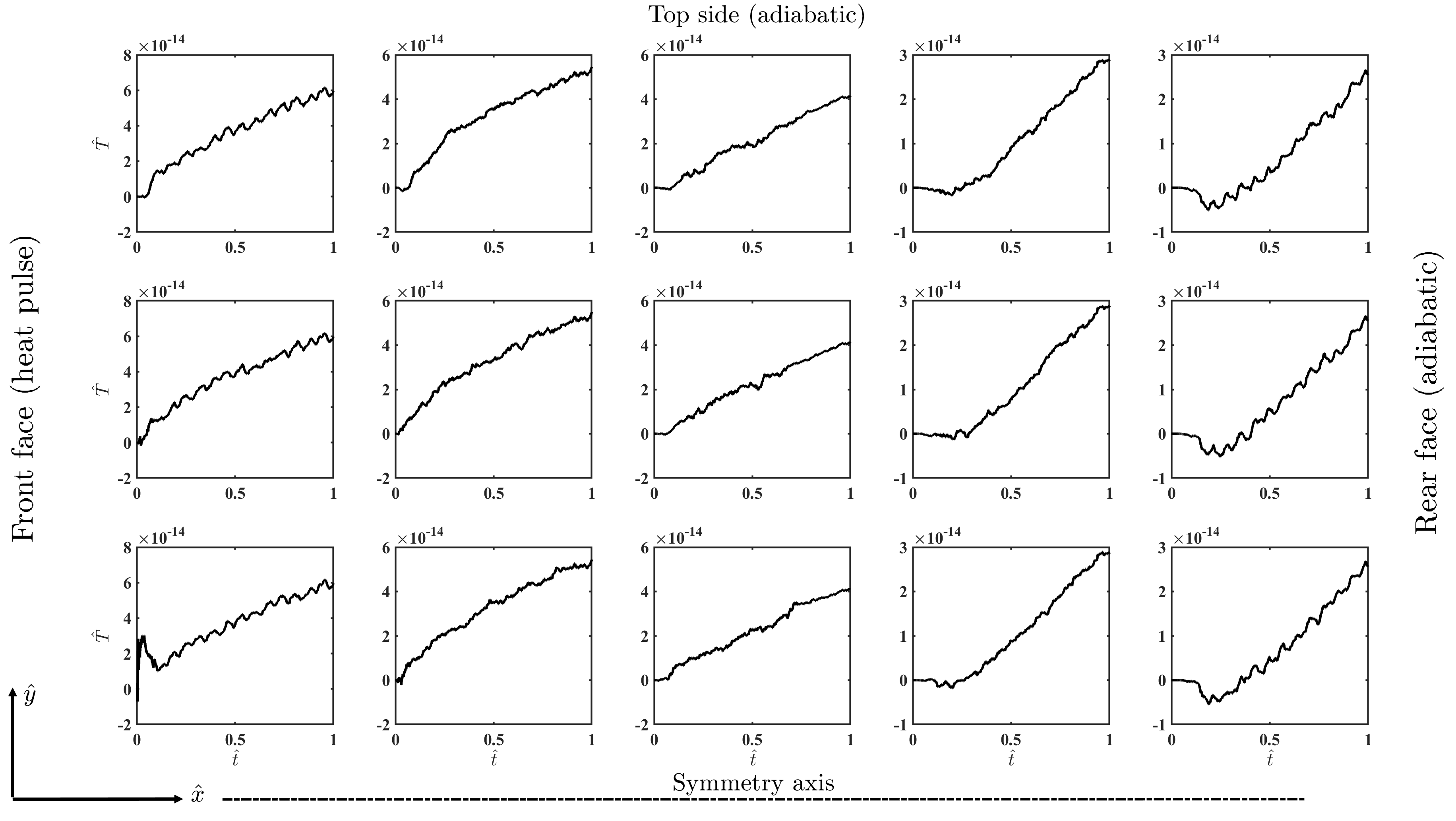}
    \caption{The difference between the Fourier and GK temperature histories when $\hat{\eta}_1=0$, and $\hat{\eta}_2=\hat \tau=0.05$.}
    \label{fig:GK1}
\end{figure}

\begin{figure}[H]
    \centering
     \includegraphics[width=1\textwidth]{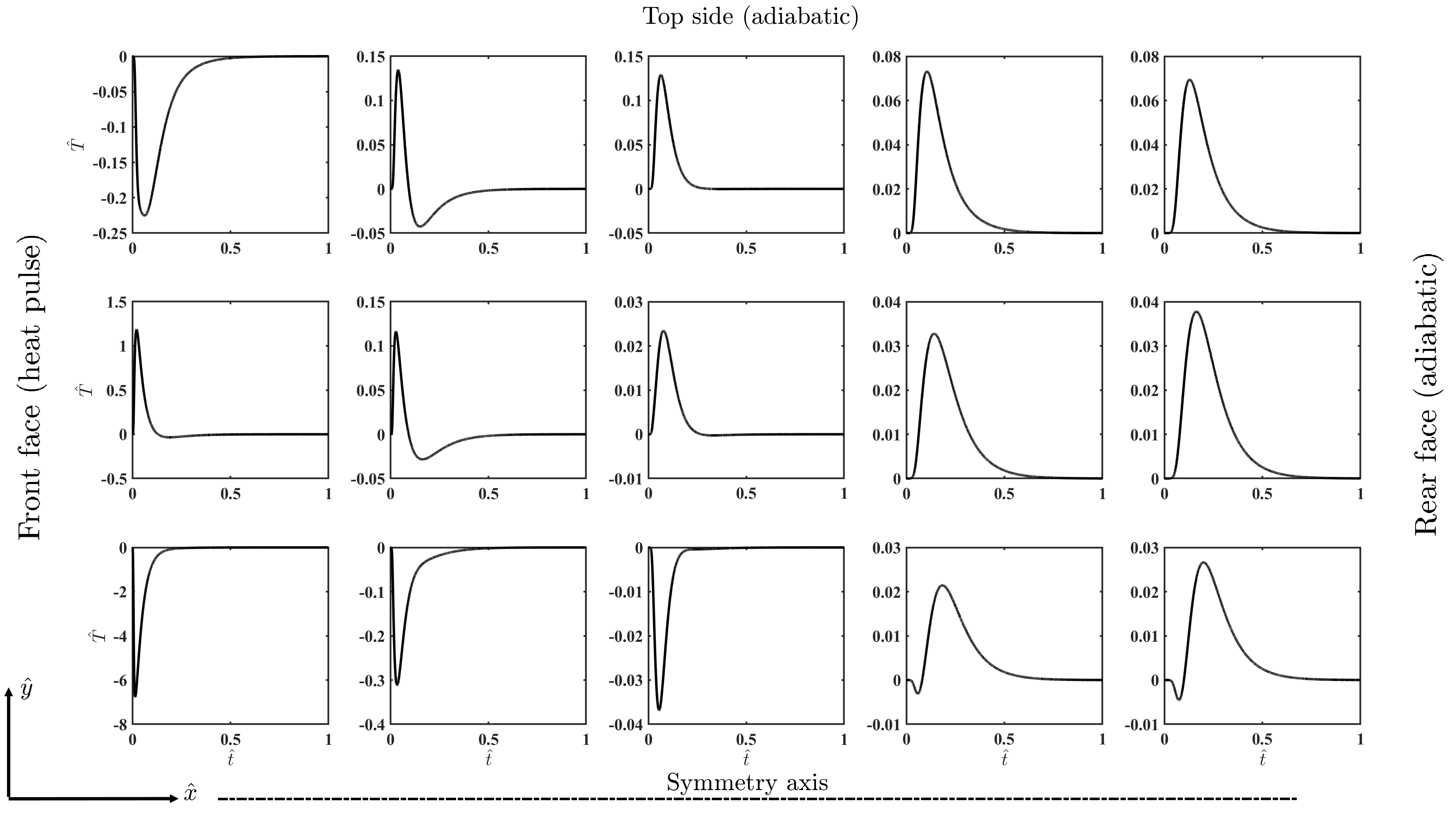}
    \caption{The difference between the Fourier and GK temperature histories when $\hat{\eta}_1=\hat \eta_2=0.025$, and $\hat \tau=0.05$.}
    \label{fig:GK2}
\end{figure}

First, we start by showing that the GK equation reproduces the Fourier solutions when $\hat{\eta}_1=0$, and $\hat{\eta}_2=\hat \tau=0.05$, Figure \ref{fig:GK1} shows the difference between the temperature fields. The observed errors are practically zero. This solution also supports our handling of $\qqQ$ on the boundary and the extrapolation method as it reproduces the Fourier solution in a particular parameter setting. In this parameter setting, the $\nabla \times \nabla \times \qqq$ term vanishes in the Eq.~\eqref{eq:GK-curl}, thus the Laplacian of $\qqq$ remains. Notable differences emerge when we keep $\hat \tau=0.05$, but $\hat \eta_1 = \hat \eta_2 = 0.025$, \ie despite that $\hat \eta_1 + \hat \eta_2 = \hat \tau$, the resonance condition is violated and the solution significantly differs from the Fourier equation, see Figure \ref{fig:GK2} for details.
Let us turn our attention to the more interesting and intriguing solutions when $\hat{\eta}_1\neq0$, and thus the curl of the heat flux field becomes meaningful.

\subsubsection*{Vorticity-free solution}
\begin{figure}[H]
    \centering
     \includegraphics[width=1\textwidth]{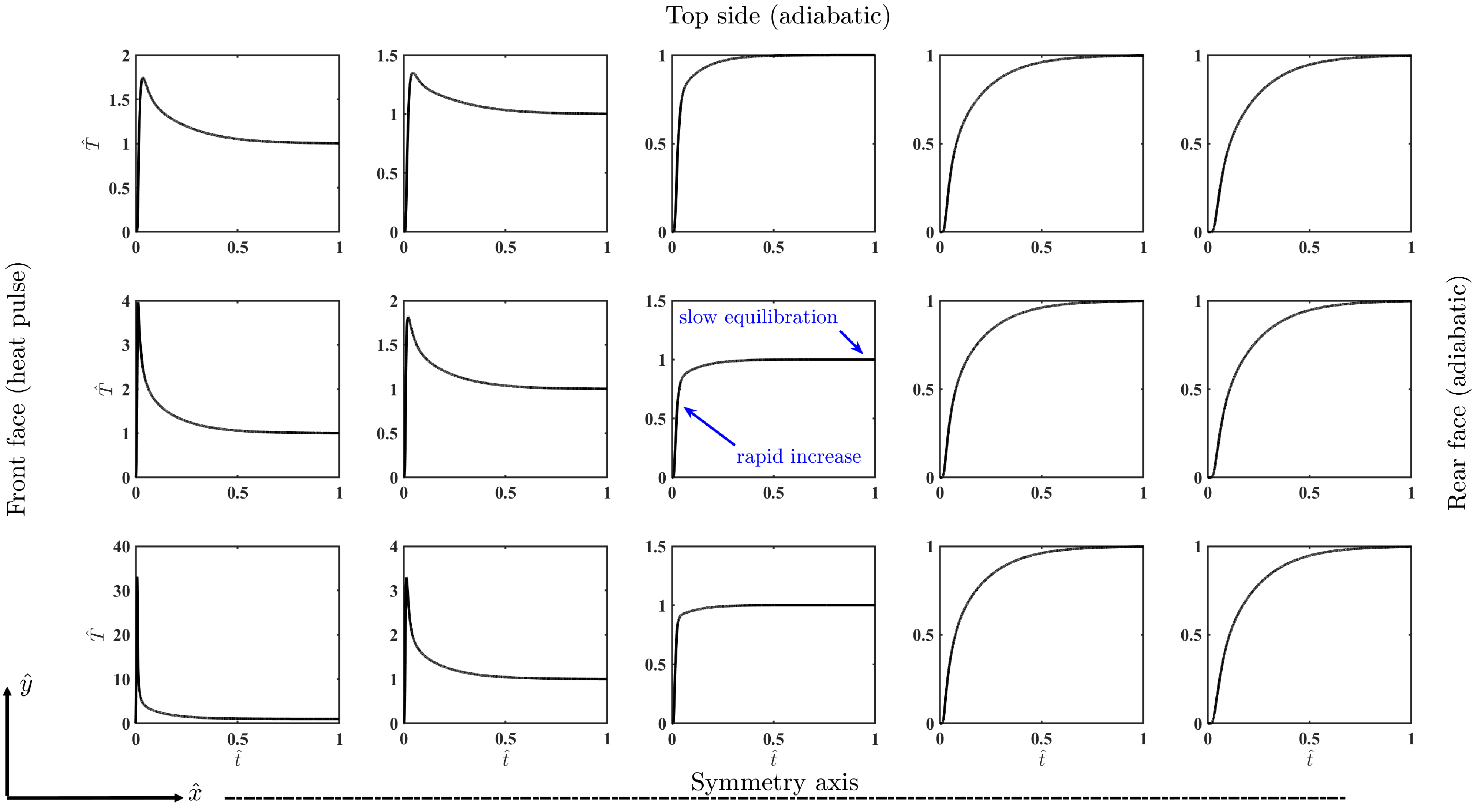}
    \caption{The GK temperature history when $\hat{\eta}_1=0$, $\hat \eta_2=0.1$, and $\hat \tau=0.05$.}
    \label{fig:GK3}
\end{figure}

\begin{figure}[H]
    \centering
     \includegraphics[width=1\textwidth]{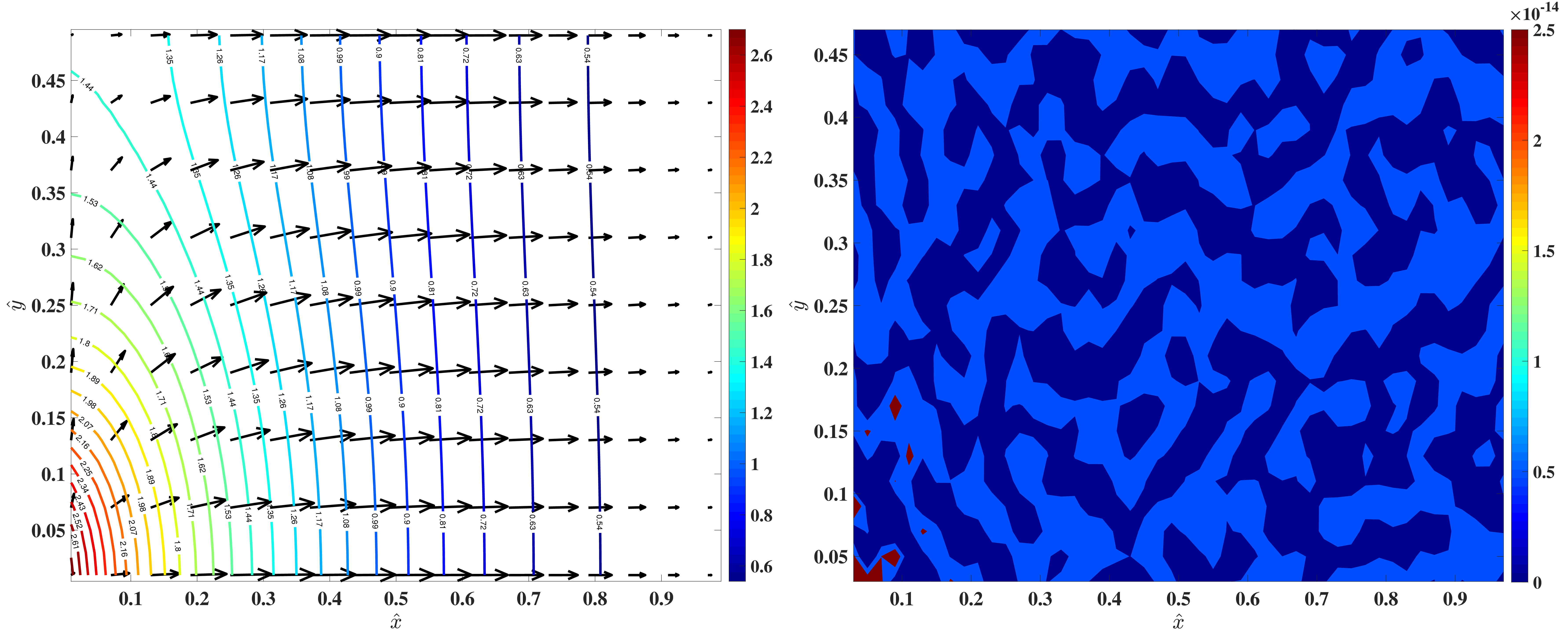}
    \caption{Left: 2D vector plot of the heat flux field when $\hat{\eta}_1=0$, $\hat \eta_2=0.1$, and $\hat \tau=0.05$ at $\hat t=0.1$. Right: the curl of the heat flux field is practically zero in accordance with the parameter settings.}
    \label{fig:GK4&5}
\end{figure}

We call the solution to be over-diffusive if $\hat \eta_1 + \hat \eta_2 > \hat \tau$. Let us start with the case when we keep $\hat \eta_1 = 0$ in order to avoid the effects of the nonzero curl of the heat flux field. It is worth comparing the characteristics of the temperature history in the middle to the rear side. In the middle, the two distinct time scales are apparent on the contrary to the rear side (see Figure \ref{fig:GK3}). In a heat pulse experiment using heterogeneous materials, such two distinct time scales are also visible, and that numerical solution reflects the size dependence of the observation \cite{FehEtal21}. This effect might not be apparent for thicker samples, and it also depends on the material properties. Figure \ref{fig:GK4&5} shows the 2D vector plot of the heat flux in which the over-diffusive behavior remains hidden; the temperature contours (isothermal lines) are slightly distorted compared to the Fourier case. Furthermore, since $\eta=0$, the curl of the heat flux field is expected to be zero, and this is reflected by Figure \ref{fig:GK4&5}, too.

\subsubsection*{Solutions with strong vorticity}
\begin{figure}[H]
    \centering
     \includegraphics[width=1\textwidth]{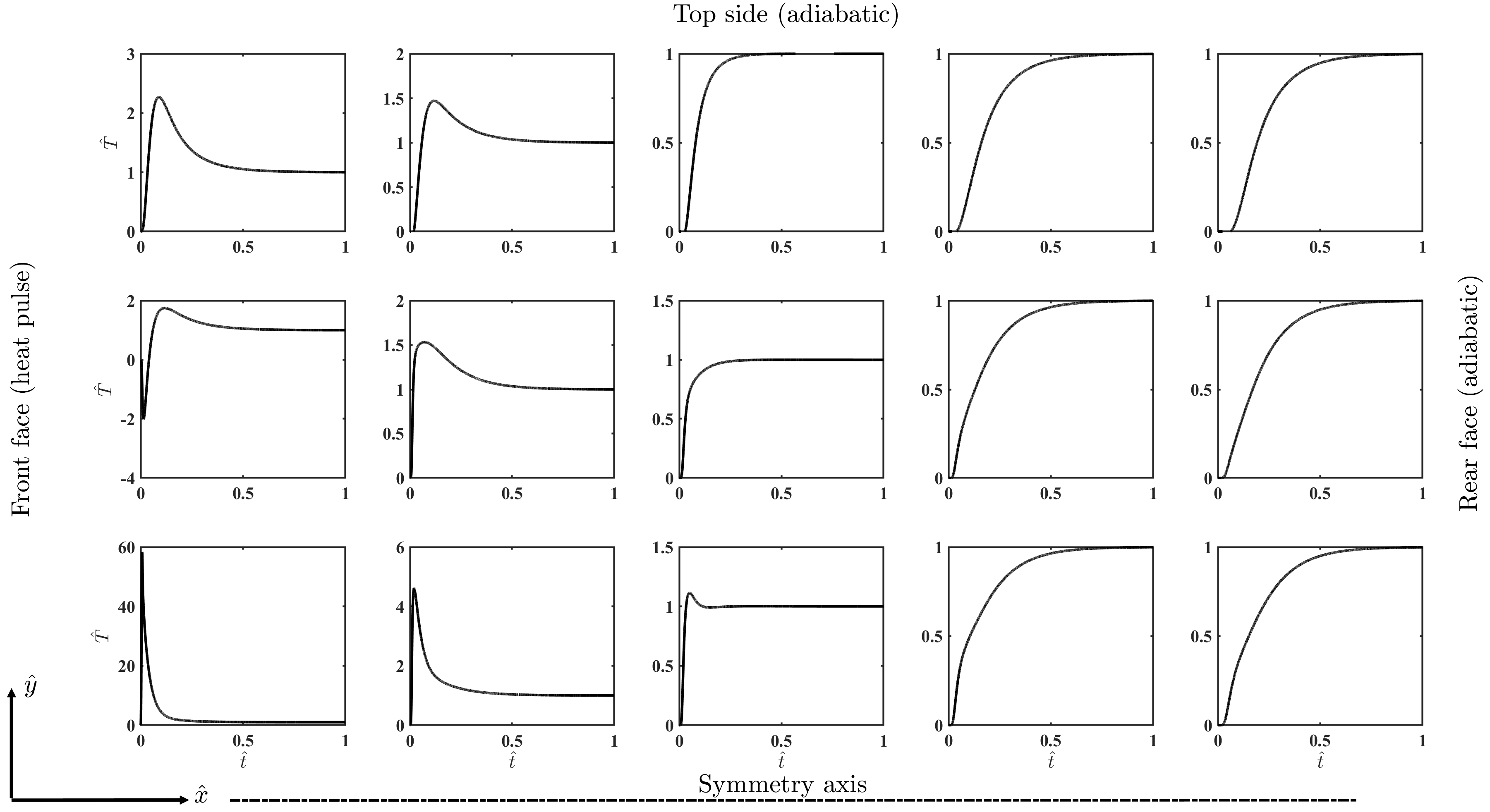}
    \caption{The GK temperature history when $\hat{\eta}_1=0.075$, $\hat \eta_2=0$, and $\hat \tau=0.05$.}
    \label{fig:GK6}
\end{figure}

\begin{figure}[H]
    \centering
     \includegraphics[width=1\textwidth]{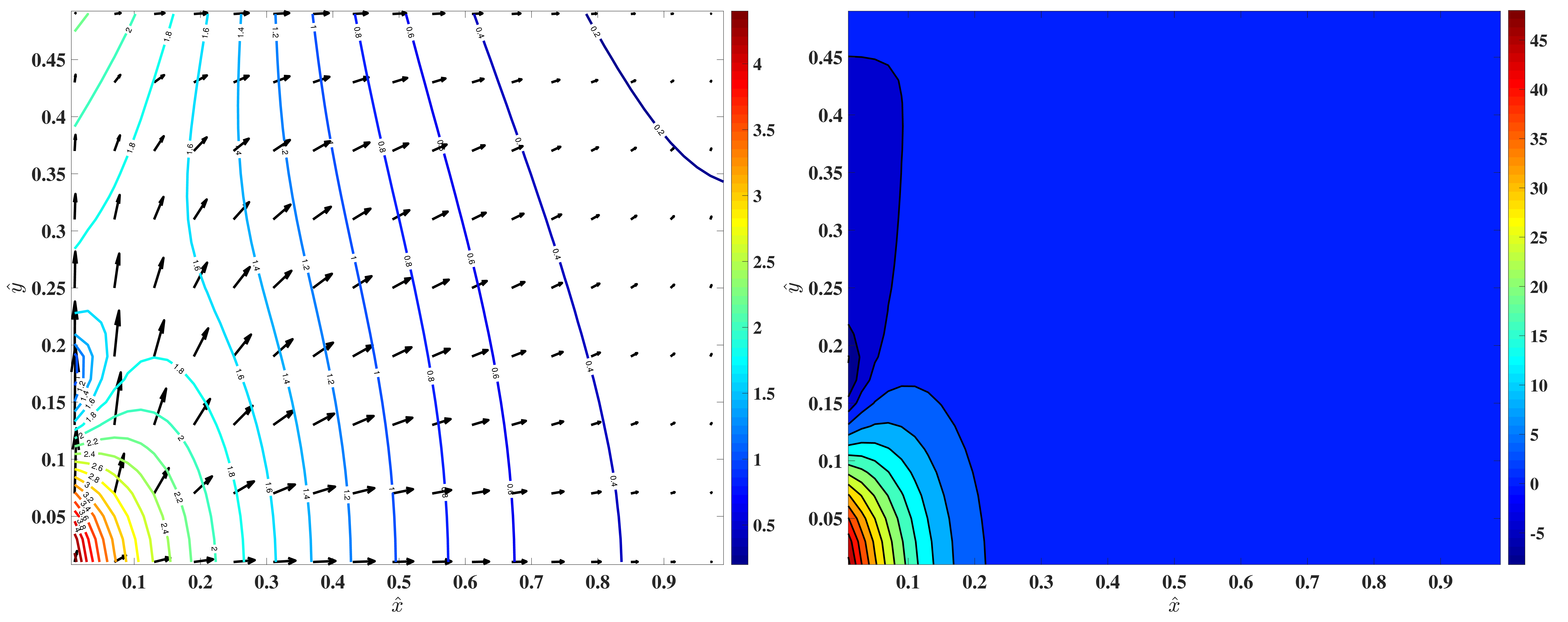}
    \caption{Left: 2D vector plot of the heat flux field when $\hat{\eta}_1=0.075$, $\hat \eta_2=0$, and $\hat \tau=0.05$ at $\hat t=0.1$. Right: the contour plot of the temperature field, highlighting the temperature-decrease effect next to the heat pulse at $\hat t=0.01$.}
    \label{fig:GK7&8}
\end{figure}

\begin{figure}[H]
    \centering
     \includegraphics[width=1\textwidth]{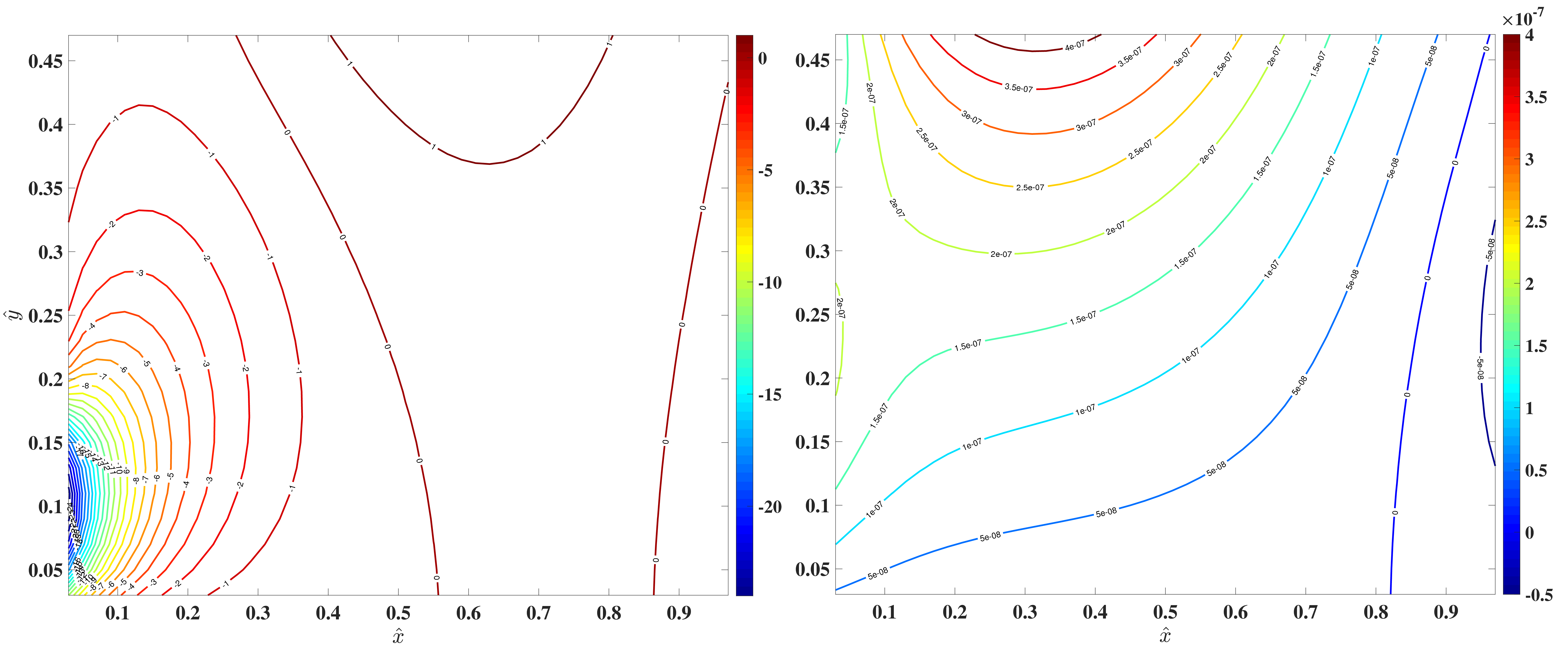}
    \caption{Left: the curl of the heat flux field when $\hat{\eta}_1=0.075$, $\hat \eta_2=0$, and $\hat \tau=0.05$ right after the heat pulse at $\hat t=0.01$. Right: the curl vanishes with time and becomes practically zero (at $\hat t=1$). }
    \label{fig:GK9&10}
\end{figure}

We turn our attention to the reverse case in which we keep $\hat \eta_2 = 0$, and now let us investigate the effects of the parameter $\hat \eta_1$, let it be $\hat \eta_1 = 0.075$. Figure \ref{fig:GK6} presents the temperature history for the given spatial points. Comparing it to the previous case, we can observe a notably different behavior. {First, near the front face, the temperature decreases due to the significant curl effects. We emphasize that this is not identical to reaching a negative absolute temperature. We wish to emphasize that the apparent negative temperature is relative to the initial temperature on contrary to the observations of Zhukovsky \cite{Zhukov16}.}
However, the temperature field can exhibit unusual evolution in such a particular parameter setting since the GK equation is based on a hydrodynamic analogy, and the curl of the heat flux field can naturally appear. Here, we particularly strengthened this effect to make it easily observable. This is only meaningful in a two- or three-dimensional setting. This temperature-decreasing effect disappears soon and is not observable for any other spatial domains. This is also depicted in Figure \ref{fig:GK7&8}. Furthermore, contrary to the previous situations, the curl of the heat flux field becomes significantly larger (Figure \ref{fig:GK9&10}).

\subsubsection*{Devating from the phonon hydrodynamic ratio}
\begin{figure}[H]
    \centering
     \includegraphics[width=0.6\textwidth]{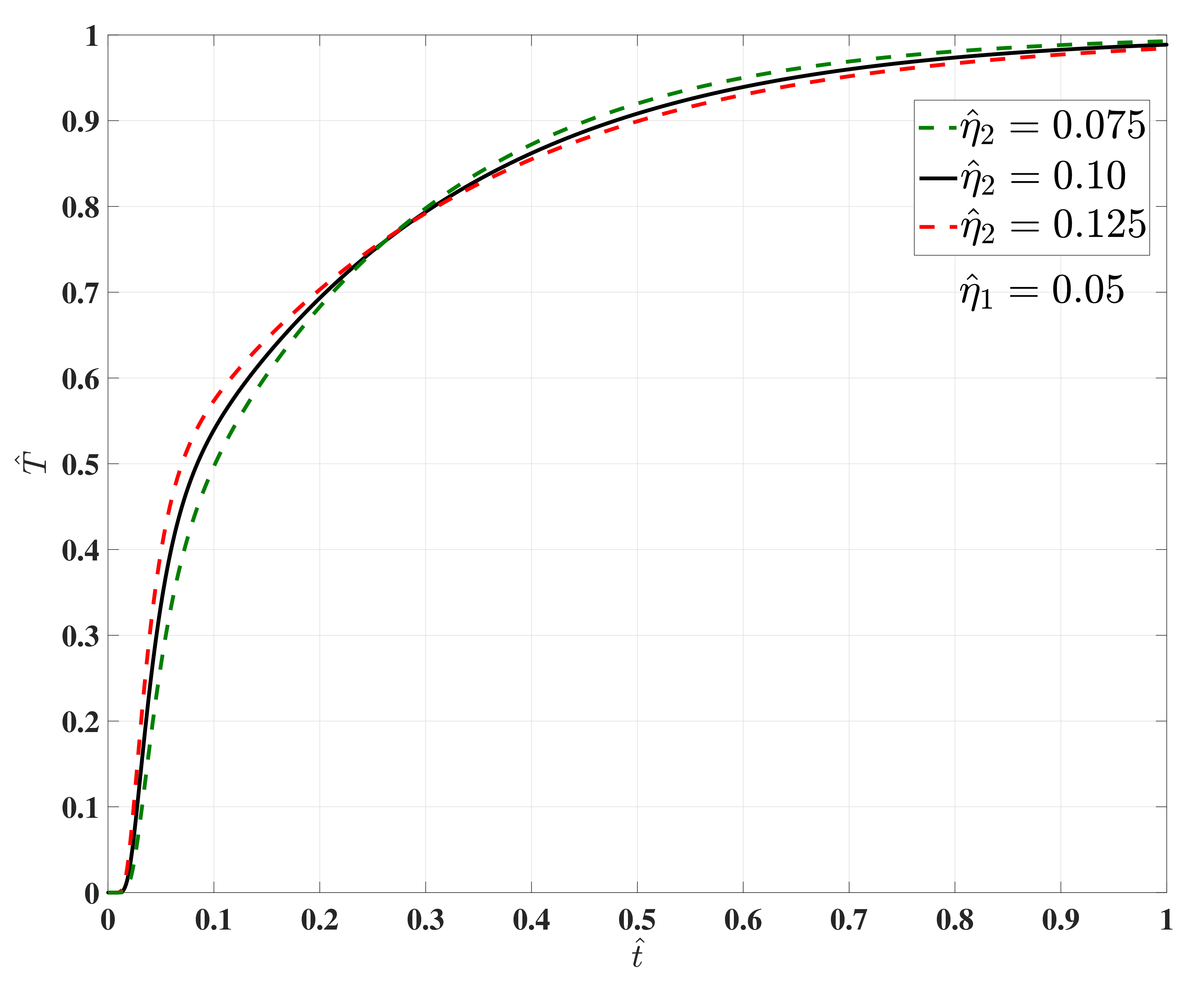}
    \caption{The GK temperature history on the rear side ($\hat x=1$, $\hat y=0.25$) when $\hat{\eta}_1=0.05$, $\hat \tau=0.05$, and $\hat \eta_2=\{0.075, 0.1, 0.125\}$, demonstrating the deviation from the phonon hydrodynamic ratio $\hat \eta_2/\hat \eta_1=2$.}
    \label{fig:GK11}
\end{figure}

\begin{figure}[H]
    \centering
     \includegraphics[width=0.6\textwidth]{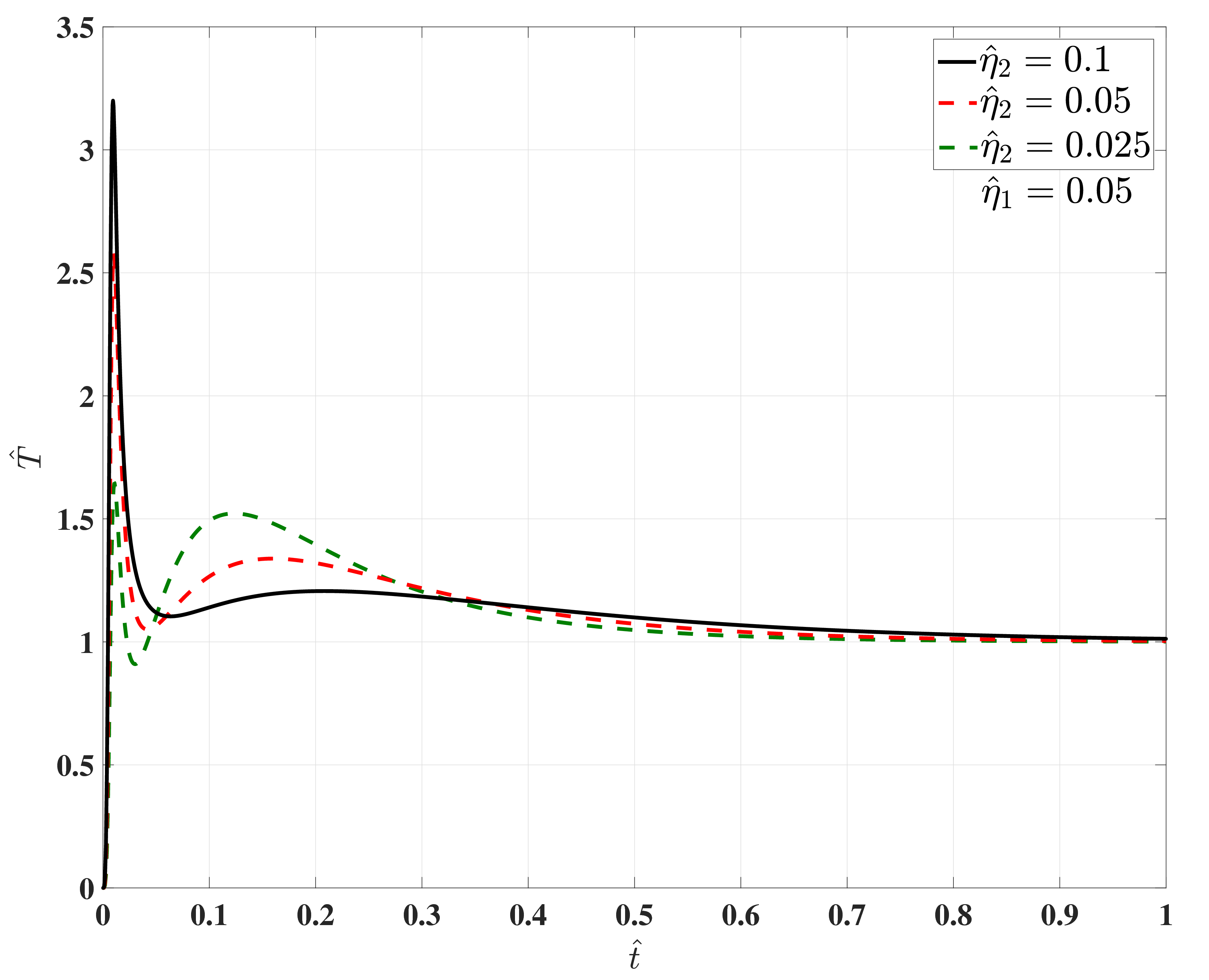}
    \caption{The GK temperature history on the front side ($\hat x=0$, $\hat y=0.25$) when $\hat{\eta}_1=0.05$, $\hat \tau=0.05$, and $\hat \eta_2=\{0.1, 0.05, 0.025\}$.}
    \label{fig:GK12}
\end{figure}

We wish to recall that the ratio $\hat \eta_2/\hat \eta_1 =2$ is fixed in a phonon hydrodynamic approach, but this does not necessarily hold in a continuum framework. In order to make this difference apparent, we provide solutions with respect to $\hat \eta_2/\hat \eta_1$. Figure \ref{fig:GK11} shows the rear side temperature history for three situations in which $\hat \eta_2/\hat \eta_1=\{1.5, 2, 2.5\}$ with fixed $\hat \eta_1=0.05$. This does not show any remarkable properties compared to the one-dimensional case, for which only the effect of $\hat \eta_1 + \hat \eta_2 $ is observable, and the increase of the over-diffusion makes the temperature signal propagation to be faster. However, if we consider the front-side temperature history in the middle (Figure \ref{fig:GK12}), then it makes more visible how the ratio of $\hat \eta_2/\hat \eta_1$ modifies the solution. Decreasing $\hat \eta_2/\hat \eta_1$ amplifies the temperature-decrease effect near the heat pulse since $\hat \eta_1$ -- the rotational part -- becomes more dominant. 

\subsubsection*{{How does the curl of heat current density behave on the boundary?}}
\begin{figure}[H]
    \centering
     \includegraphics[width=1\textwidth]{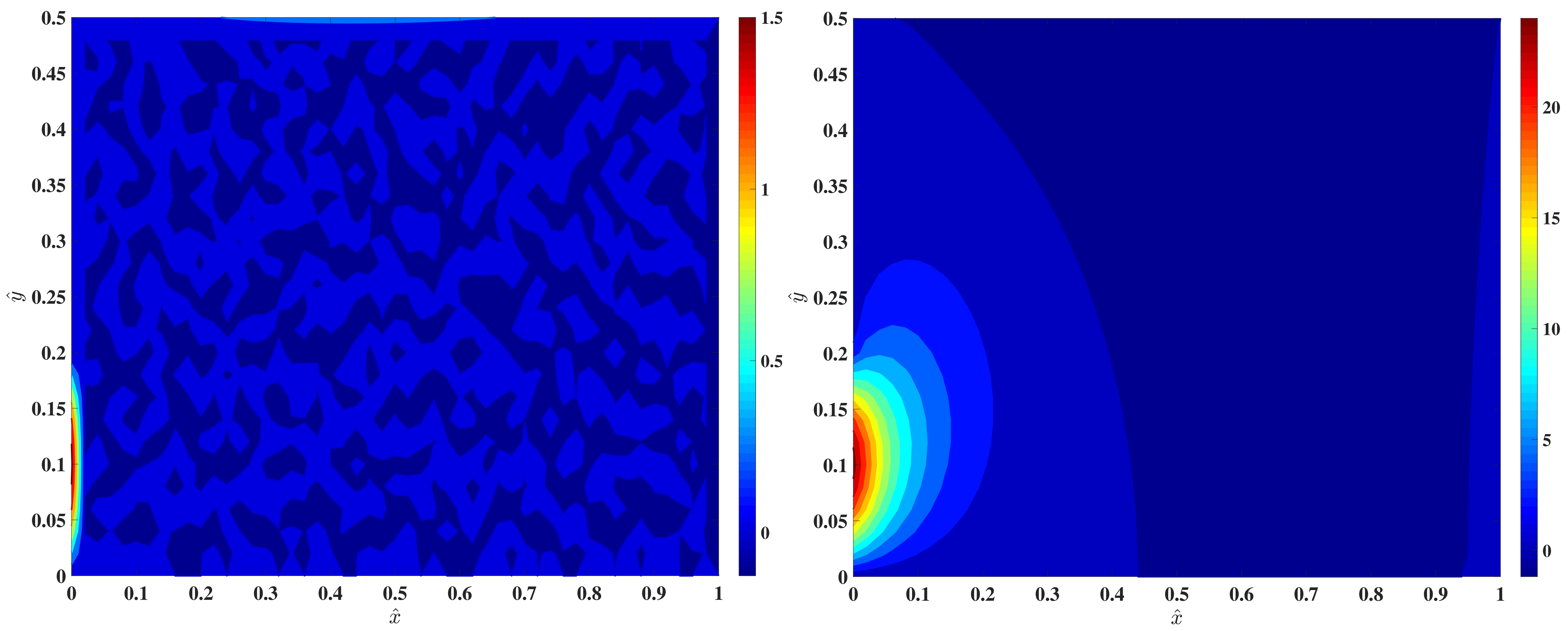}
    \caption{{Left: $ \left( \qqq \times \nabla \right)_z $ without the rotational term at $\hat t=0.1$. Right: $ -\left( \qqq \times \nabla \right)_z $ with the rotational term at $\hat t=0.1$. The sign is changed in order to have colors correctly emphasizing the differences in magnitudes.}}
    \label{fig:GK13}
\end{figure}

\begin{figure}[H]
	\centering
	\includegraphics[width=0.5\textwidth]{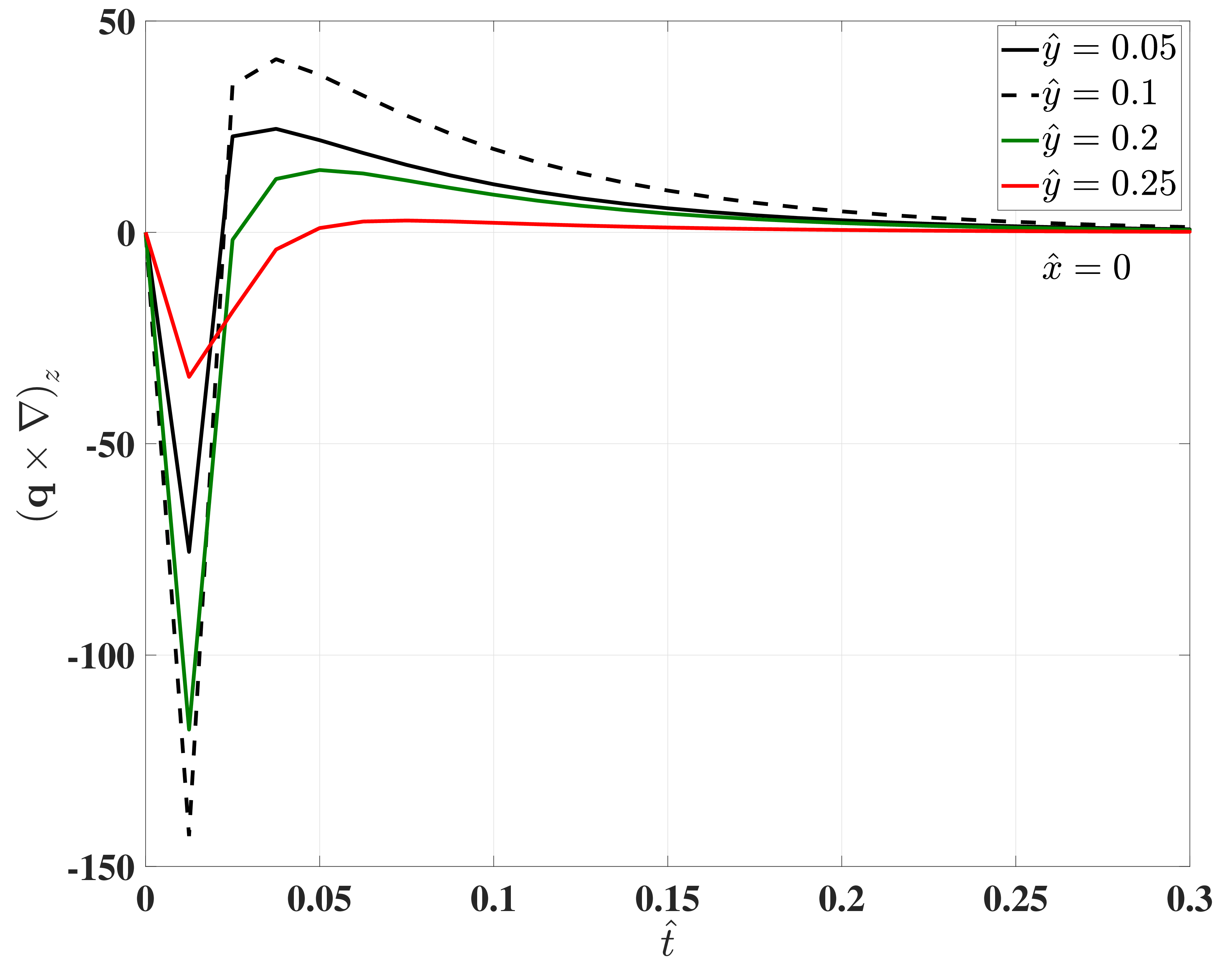}
	\caption{The time evolution of  $ \left( \mathbf{q} \times \nabla  \right)_z$ including the rotational term on the front face, indicating that the vorticities are the strongest at the heat pulse, and quickly attenuate after $\hat y=0.2$. Furthermore, shortly after the heat pulse, the exponential decay is apparent in agreement with Eq.~\eqref{eq:gk0}.}
	\label{fig:GK14}
\end{figure}

Previously, we introduced the auxiliary field $\qqQ$ in order to ease the discretization of the second-order spatial derivatives and make it easier to realize the boundary conditions properly. Since the diagonal elements of $\qqQ$ are inside the spatial domain, they are not directly related to the boundary conditions. However, the off-diagonals are not independent of the $\qqq$-boundary. We introduced an extrapolation from the bulk in order to avoid the definition of incompatible boundary data for the unknown off-diagonals, and thus avoiding the introduction of any artificial distortion. Now let us depict the difference between the $Q_{xy}$ and $Q_{yx}$ components in two cases. {In fact, this difference is the only component of the curl of the in-plane heat current density.} In the first one, $\hat \eta_1=0$ is considered, hence, the rotational term is zero ($\hat \eta_2 = \hat \tau=0.05$). In the second case, $\hat \eta_1=\hat \eta_2 = 0.05$, and notable differences are expected. Figure \ref{fig:GK13} presents their difference, highlighting that $\hat \eta_1$ indeed introduces significant changes in the evolution of off-diagonals of $\qqQ$, especially near the boundaries, but also notably affecting the bulk behavior. Figure \ref{fig:GK14} shows the time evolution in agreement with Eq.~\eqref{eq:gk0}, presenting an exponential decay in time.

\section{Discussion and summary}

In the present paper, we revisited the continuum thermodynamic background of the Guyer--Krumhansl equation. With distinguishing between the internal variable and its entropy-conjugate, we revealed the need for a further constitutive relationship, which connect these two together. This additional possibility remains hidden when a particular extension for the entropy density is assumed, especially when the internal variable is immediately identified with the heat flux density. Although we do not investigate its outcomes in greater detail here, we note that it poses further potential and could have non-trivial consequences. 

Furthermore, we applied a staggered discretization approach to numerically solve a two-dimensional setting including a space and time-dependent heat flux boundary condition. The discretization is eased by introducing a second-order tensory, $\qqQ$ as an auxiliary quantity, and $\qqQ$ is also helpful in the proper realization of boundary conditions. We discussed that $\qqQ$ is not independent of the given $\qqq$ boundary condition, however, not all components of $\qqQ$ can be found immediately. In order to avoid artificial and unnecessary assumptions, we proposed to use a quadratic Lagrangian extrapolation based on the bulk points to update the unknown $\qqQ$ components on the boundary. That approach successfully reproduced the solutions of the Fourier equation applying the resonance condition ($\hat \eta_1=0$, \ie with vanishing rotational term).

Additionally, the continuum background of the GK equation allowed to adjust the parameters in a range, which is beyond the validity of phonon hydrodynamics. In a case when the rotational terms dominate the heat flux density time evolution, we could observe that the temperature can be significantly decreased, even under the initial temperature. However, we want to strongly emphasize that this phenomenon is not identical to obtain negative temperature, and occurs only locally for a brief period. This is a characteristic outcome of the whirling heat current density. 

Our analysis also revealed that when the GK equation is used as an effective description of heterogeneous materials, $\hat \eta_1=0$ is a necessary choice in order to keep the possibility for Fourier resonance, and avoid solutions being characteristic only for the hydrodynamic domain, \eg for low-temperature or nanoscale problems. However, the continuum background always offers the possibility to inherit the coefficients from phonon hydrodynamics, and therefore our numerical approach and findings can be useful even in these situations.

\section{Acknowledgement}
The authors express their gratitude to Péter Ván for his useful suggestions. The research of C.F.M. has been carried out under the auspices
of GNFM (National Group of Mathematical-Physics) of INdAM (National Institute of
Advanced Mathematics), through the grant ‘Progetto Giovani’ CUPE53C22001930001 for
financial support. Project no.~TKP-6-6/PALY-2021 has been implemented with the support provided by the Ministry of Culture and Innovation of Hungary from the National Research, Development and Innovation Fund, financed under the TKP2021-NVA funding scheme. The research was funded by the Sustainable Development and Technologies National Programme of the Hungarian Academy of Sciences (FFT NP FTA). This work was partially supported in part by the Hungarian Scientific Research Fund under Grant agreement FK 134277.


\end{document}